\documentclass[acmsmall,screen]{acmart}

\usepackage{fancyvrb}
\usepackage{xurl}
\usepackage{booktabs}
\usepackage{url}
\usepackage{amsmath,amsfonts}
\usepackage{algorithmic}
\usepackage{graphicx}
\usepackage{textcomp}
\usepackage{xspace}
\usepackage{subcaption}
\usepackage{tabularx}
\usepackage{amsmath}
\usepackage{balance}
\usepackage{framed}
\usepackage{chngpage}
\usepackage[many]{tcolorbox}
\usepackage{pgfplots}
\usepackage{siunitx}
\usepackage{balance}
\usepackage{enumitem}
\usepackage{multirow}

\usepackage{hyperref}

\usepackage{graphicx}
\usepackage{subcaption}

\newcommand{\am}{agent context files\xspace}
\newcommand{\Am}{Agent context files\xspace}
\newcommand{\AM}{Agent Context Files\xspace}
\newcommand{\file}{agent context file\xspace}
\newcommand{\files}{agent context files\xspace}

\newcommand{\Files}{Agent context files\xspace}

\newcommand{\ACMs}{ACFs\xspace}
\newcommand{\numfiles}{2,303\xspace}
\newcommand{\numrepos}{1,925\xspace}

\newcommand{\numlabelingfiles}{332\xspace}

\newcommand{\claude}{Claude Code\xspace}
\newcommand{\copilot}{GitHub Copilot\xspace}
\newcommand{\codex}{OpenAI Codex\xspace}

\newcommand{\claudemd}{\texttt{CLAUDE.md}\xspace}
\newcommand{\copilotmd}{\texttt{copilot-instructions.md}\xspace}
\newcommand{\agentsmd}{\texttt{AGENTS.md}\xspace}

\newcommand{\claudenumfiles}{922\xspace}
\newcommand{\claudenumrepos}{922\xspace}
\newcommand{\claudenumcommits}{5,655\xspace}
\newcommand{\codexnumfiles}{694\xspace}
\newcommand{\codexnumrepos}{694\xspace}
\newcommand{\codexnumcommits}{2,767\xspace}
\newcommand{\copilotnumfiles}{687\xspace}

\newcommand{\copilotnumcommits}{2,237\xspace}

\newcommand{\rqa}{$RQ_1$}
\newcommand{\rqc}{$RQ_2$}
\newcommand{\rqd}{$RQ_3$}
\newcommand{\rqe}{$RQ_4$}
\newcommand{\rqaa}{What are the characteristics of \am?}

\newcommand{\rqcc}{How often do developers maintain \am?}
\newcommand{\rqdd}{What instructions are included in \am?}
\newcommand{\rqee}{To what extent can instructions in \am be classified automatically?}

\newcommand{\rqA}{\rqa: \rqaa}

\newcommand{\rqC}{\rqc: \rqcc}
\newcommand{\rqD}{\rqd: \rqdd}
\newcommand{\rqE}{\rqe: \rqee}

\newcommand{\lblGenSysOverview}{\texttt{System Overview}\xspace}
\newcommand{\cntGenSysOverview}{59.0}

\newcommand{\lblClaudeAIIntegration}{\texttt{AI Integration}\xspace}
\newcommand{\cntClaudeAIIntegration}{24.1}

\newcommand{\lblDocRefs}{\texttt{Documentation and References}\xspace}
\newcommand{\cntDocRefs}{26.8}

\newcommand{\lblArchitecture}{\texttt{Architecture}\xspace}
\newcommand{\cntArchitecture}{68.1}

\newcommand{\lblImplDetails}{\texttt{Implementation Details}\xspace}
\newcommand{\cntImplDetails}{70.8}

\newcommand{\lblBuildRun}{\texttt{Build and Run}\xspace}
\newcommand{\cntBuildRun}{63.0}

\newcommand{\lblTest}{\texttt{Testing}\xspace}
\newcommand{\cntTest}{75.9}

\newcommand{\cntConfigEnv}{38.9}

\newcommand{\lblDeployOps}{\texttt{DevOps}\xspace}
\newcommand{\cntDeployOps}{18.4}

\newcommand{\lblMaintenance}{\texttt{Maintenance}\xspace}
\newcommand{\cntMaintenance}{44.6}

\newcommand{\lblProjMgmt}{\texttt{Project Management}\xspace}
\newcommand{\cntProjMgmt}{5.4}

\newcommand{\lblDevProcess}{\texttt{Development Process}\xspace}
\newcommand{\cntDevProcess}{65.1}

\newcommand{\lblPerformance}{\texttt{Performance}\xspace}
\newcommand{\cntPerformance}{14.5}

\newcommand{\lblSecurity}{\texttt{Security}\xspace}
\newcommand{\cntSecurity}{14.8}

\newcommand{\lblUIUX}{\texttt{UI/UX}\xspace}
\newcommand{\cntUIUX}{8.7}

\newcommand{\lblDebugging}{\texttt{Debugging}\xspace}
\newcommand{\cntDebugging}{25.3}

\newcommand{\addinstline}{\texttt{add:instructions/line}\xspace}
\newcommand{\addinstsect}{\texttt{add:instructions/section}\xspace}
\newcommand{\delinstline}{\texttt{del:instructions/line}\xspace}
\newcommand{\delinstsect}{\texttt{del:instructions/section}\xspace}
\newcommand{\maintenance}{\texttt{maintenance}\xspace}
\newcommand{\fixinstr}{\texttt{fix:instructions}\xspace}

\newcommand{\firstLabels}{2,227\xspace}

\newcommand{\agreementRate}{80.3\%\xspace}
\newcommand{\conflictingLabels}{438\xspace}
\newcommand{\totalLabels}{2,069\xspace}

\definecolor{darkgreen}{rgb}{0, 0.5, 0} 
\definecolor{whitesmoke}{rgb}{0.99, 0.99, 0.99} 

\def\Underline{\setbox0\hbox\bgroup\let\\\endUnderline}
\def\endUnderline{\vphantom{y}\egroup\smash{\underline{\box0}}\\}
\def\|{\verb|}

\newcommand{\ie}{\textit{i.e.,}\xspace}
\newcommand{\eg}{\textit{e.g.,}\xspace}

\newcounter{findings_no}

\usepackage{listings}
\definecolor{backcolour}{rgb}{0.95,0.95,0.92}
\lstdefinelanguage{diff}{
  morecomment=**[f][\color{red}]{-},         
  morecomment=**[f][\color{darkgreen}]{+},       
  moredelim=**[is][\bfseries]{@@}{@@},
}
\definecolor{backcolour}{rgb}{0.95,0.95,0.92}
\lstdefinelanguage{commit}{ 
  breakindent = 0pt,
  numbers=none,
  backgroundcolor=\color{white},
  frame=single,
  xleftmargin=3.5em,
  numbersep=0em,
  xrightmargin=1.5em,
}

\usepackage[many]{tcolorbox}  
\tcbuselibrary{listings,breakable}
\definecolor{main}{HTML}{D0D3D4}    
\definecolor{sub}{HTML}{D0D3D4}     
\newtcolorbox{dbox}{
    left=2pt,right=2pt,top=2pt,bottom=2pt,
    enhanced, 
    boxrule = 0pt,
    enlarge top by=5pt,
    enlarge bottom by=3pt,
  }
\tcbset{
    sharp corners,
    before skip = 0.1cm,    
    after skip = 0.01cm      
}

\def\summarybox#1#2{
\medskip
\begin{tcolorbox}[
  enhanced,
  title=#1,
  colframe=darkgray,
]
    #2
\end{tcolorbox}
}

\lstdefinestyle{academicstyle}{
    basicstyle=\ttfamily\footnotesize,
    breaklines=true,
    frame=single,
    xleftmargin=2em,
    xrightmargin=2em,
    aboveskip=0.5em,
    belowskip=2em,
    abovecaptionskip=0em,      
    belowcaptionskip=0.5em,      
    columns=flexible,
    keepspaces=true,
    language={},  
}

\usepackage{tcolorbox}
\tcbuselibrary{skins,breakable}
\newtcolorbox{reviewcomment}[1][Comment]{  
    enhanced,
    frame hidden,
    borderline west={4pt}{0pt}{gray!50},
    colback=gray!10,
    boxrule=0pt,
    sharp corners,
    breakable,
    before upper={\parindent0pt\textbf{#1:}\par\vspace{0.3em}},
    fontupper=\itshape,
    left=10pt,
    right=10pt,
    top=5pt,
    bottom=5pt,
    before skip=1.5\baselineskip
}

\AtBeginDocument{%
  \providecommand\BibTeX{{%
    \normalfont B\kern-0.5em{\scshape i\kern-0.25em b}\kern-0.8em\TeX}}}

\setcopyright{acmlicensed}
\copyrightyear{2025}
\acmYear{2025}

\acmISBN{978-1-4503-XXXX-X/18/06}

\begin{document}

\title{Agent READMEs: An Empirical Study of Context Files for Agentic Coding}

\author{Worawalan Chatlatanagulchai}
\affiliation{%
  \institution{Faculty of Engineering, Kasetsart University}
  \city{Bangkok}
  \country{Thailand}
}
\affiliation{%
  \institution{Nara Institute of Science and Technology}
  \city{Ikoma}
  \country{Japan}
}

\author{Hao Li}
\affiliation{%
  \institution{Queen's University}
  \city{Kingston}
  \country{Canada}
}

\author{Yutaro Kashiwa}
\affiliation{%
  \institution{Nara Institute of Science and Technology}
  \city{Ikoma}
  \country{Japan}
}
\email{yutaro.kashiwa@is.naist.jp}

\author{Brittany Reid}
\affiliation{%
  \institution{Nara Institute of Science and Technology}
  \city{Ikoma}
  \country{Japan}
}

\author{Kundjanasith Thonglek}
\affiliation{%
  \institution{Faculty of Engineering, Kasetsart University}
  \city{Bangkok}
  \country{Thailand}
}



\author{Pattara Leelaprute}
\affiliation{%
  \institution{Faculty of Engineering, Kasetsart University}
  \city{Bangkok}
  \country{Thailand}
}

\author{Arnon Rungsawang}
\affiliation{%
  \institution{Faculty of Engineering, Kasetsart University}
  \city{Bangkok}
  \country{Thailand}
}
\author{Bundit Manaskasemsak}
\affiliation{%
  \institution{Faculty of Engineering, Kasetsart University}
  \city{Bangkok}
  \country{Thailand}
}

\author{Bram Adams}
\affiliation{%
  \institution{Queen's University}
  \city{Kingston}
  \country{Canada}
}

\author{Ahmed E. Hassan}
\affiliation{%
  \institution{Queen's University}
  \city{Kingston}
  \country{Canada}
}

\author{Hajimu Iida}
\affiliation{%
  \institution{Nara Institute of Science and Technology}
  \city{Ikoma}
  \country{Japan}
}

\renewcommand{\shortauthors}{Chatlatanagulchai et al.}

\begin{abstract}
Agentic coding tools receive goals written in natural language, break them down into specific tasks, and write or execute code with minimal human intervention. Central to this process are agent context files (e.g., AGENTS.md and CLAUDE.md) that provide persistent, project-level instructions. In this paper, we conduct the first large-scale empirical study of 2,303 agent context files from 1,925 repositories to characterize their structure, maintenance, and content. We find that these files are not static documentation but complex, difficult-to-read artifacts that evolve like configuration code through frequent, small additions. Our content analysis of 16 instruction types shows that developers prioritize functional context, such as test procedures (\cntTest\%), implementation details (\cntImplDetails\%), and architecture (\cntArchitecture\%). We also identify a significant gap: non-functional requirements such as security (\cntSecurity\%) and performance (\cntPerformance\%) are rarely specified. These findings indicate that while developers use context files to make agents functional, they provide few guardrails to ensure that agent-written code is secure or performant, highlighting the need for improved tools and practices.

\end{abstract}



\maketitle

\section{Introduction}

\begin{flushright}
  \itshape
  ``The hottest new programming language is English.''\\
  --- Andrej Karpathy (Founding member of OpenAI)
\end{flushright}
\smallskip
Large Language Models~(LLMs) are transforming software development from editing code by hand to instructing Artificial Intelligence (AI) agents in natural language~\cite{hassan2025agenticse, li2025aidev}.
This novel approach, termed \textit{Agentic Coding}, interprets natural language goals, decomposes them into subtasks, and autonomously plans and writes code with minimal human intervention. However, the autonomy of these agents hinges on a critical, often overlooked factor: \textit{context}. For an agent to work for a software project, adhere to architectural patterns, and follow team conventions, it requires not only access to source code but also explicit guidance. The effectiveness and safety of this workflow depend not only on model quality but also on how projects communicate their architecture, constraints, and conventions to these agents.

To address this, a new class of software artifact has emerged: \textit{\am}. These are specialized files (\eg \agentsmd) that work like a README for agents\footnote{\url{https://agents.md}} and define how the agent should behave within a given software project. These documents specify project-specific knowledge, architectural overviews, build and test commands, coding conventions, and explicit rules for interacting with tools and external services. They define the agent's role, describe the system architecture, record build and test commands, and encode coding standards and operational rules. Modern coding tools such as \claude, \codex, and \copilot load these files at the start of a session and use them as persistent context when planning and executing changes. In practice, the quality of these \am largely determines how well an agent can understand a repository, follow team practices, and perform non-trivial maintenance tasks.

Prior work on LLM-assisted software engineering has focused on model capabilities, interaction patterns, and immediate prompts, for example, code generation~\cite{Jinetal2024, hai2025impactscontextsrepositorylevelcode}, 
and the impact of natural language inputs on generated code quality~\cite{DBLP:conf/icse/MastropaoloPGCSOB23}, 
automatic program repair~\cite{zhang2024autocoderover, Xiaetal2022, Bouzeniaetal2024}, and multi-agent collaboration~\cite{Zhangetal_12024}. Researchers have also begun to study context engineering~\cite{Shinetal2023, hai2025impactscontextsrepositorylevelcode} and prompt management in software repositories~\cite{li2025promptgithub, Shrivastavaetal2022}. However, the concrete artifacts that teams already use in the wild to steer coding agents (namely \am) remain underexplored. Official tool documentation gives only high-level guidance (\eg ``describe architecture and workflows''), leaving developers to design these \am by trial and error. We currently lack basic empirical evidence about what these files look like, how they evolve over time, and what instructions they actually contain.


To bridge this gap, we conduct a large-scale empirical study of \am. Building on our prior work~\citep{chatlatanagulchai2025agenticmanifests} which studied 253 context files for \claude, we extend the scope to \numfiles context files from \numrepos repositories across three agentic coding tools: \claude, \codex, and \copilot. We aim to uncover the structural patterns, maintenance habits, and instructional strategies that developers currently employ. Our study focuses on these research questions~(RQs):

\begin{itemize}[leftmargin=1em]
    \item[] \textbf{\rqa: \rqaa}
    Understanding readability and structure is essential for creating maintainable agent configurations. \Am are generally long and difficult to read, yet follow a consistent shallow hierarchy with a single top-level heading and most content organized under H2 and H3 sections.
\smallskip
    \item[] \textbf{\rqc: \rqcc} 
    It is unclear whether \am behave like static documentation or like evolving configuration in the codebase. \Am are actively maintained in short bursts, evolving through incremental additions rather than deletions, behaving as living configuration artifacts rather than static documents.
\smallskip
    \item[] \textbf{\rqd: \rqdd} 
    Identifying what context developers prioritize reveals the current focus and blind spots of agentic coding in real-world projects. Instructions are heavily skewed toward functional operations (\eg \lblBuildRun, \lblImplDetails), while critical non-functional requirements like \lblSecurity and \lblPerformance are rare.
\smallskip
    \item[] \textbf{\rqe: \rqee}
    Manual analysis of \am' content is unscalable, making it critical to understand if this classification can be automated for future monitoring. Automatic classification is highly effective (0.79 F1-score) for concrete functional topics (\eg \lblTest and \lblArchitecture), but it struggles with abstract or nuanced topics (\eg \lblMaintenance).
\end{itemize}

\begin{figure}[t]
    \centering
    \includegraphics[width=0.7\linewidth]{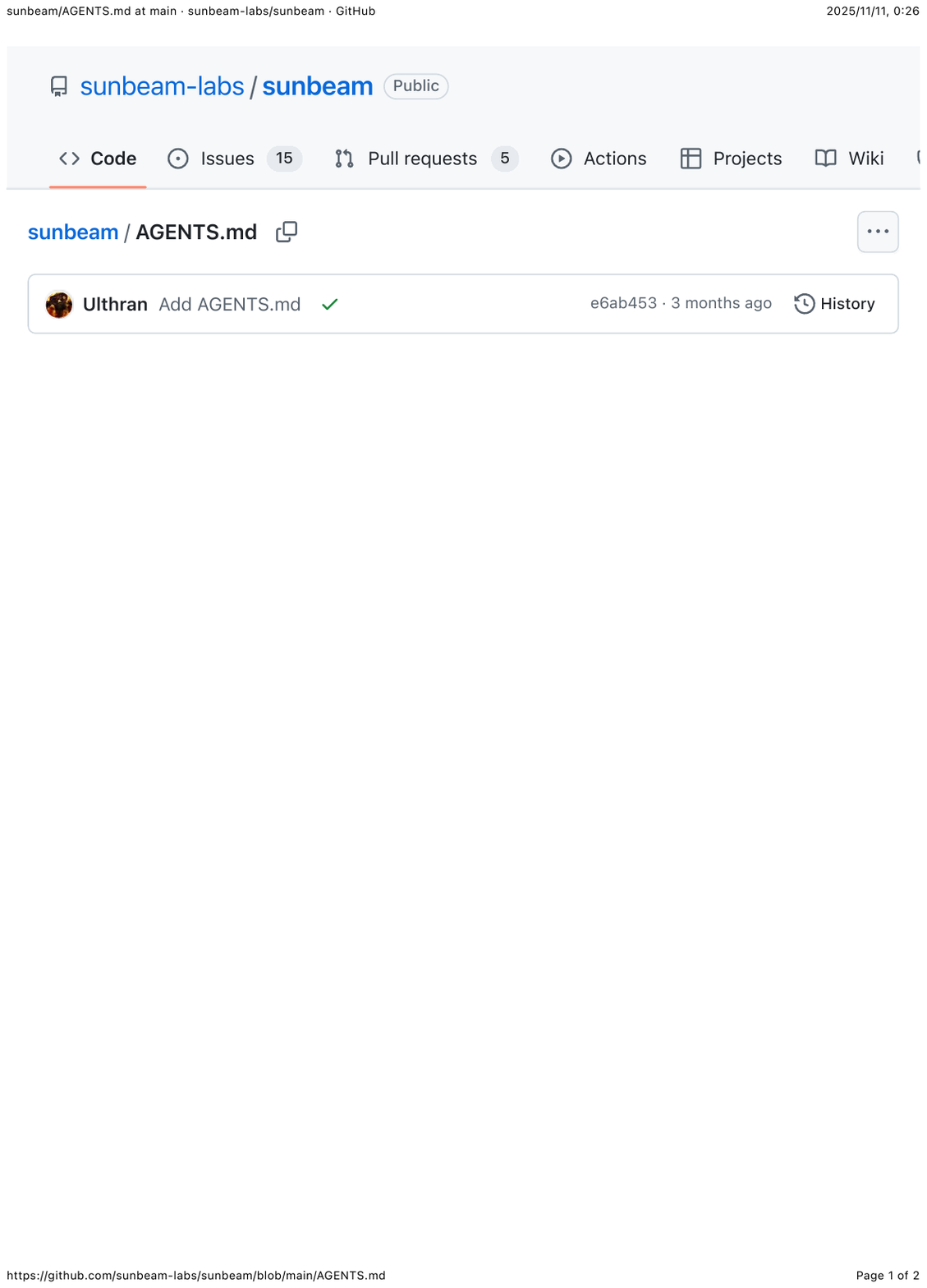}
    \caption{Example of \am}
    \label{fig:example1}
\end{figure}

\smallskip
\textbf{Replication Package.} To facilitate replication and further studies, we provide the data used in our replication package.\footnote{\url{https://github.com/woraamy/Agent-Context-File-Analysis}} 

\smallskip
\textbf{Paper Structure.} The remainder of this paper is organized as follows. Section~\ref{sec:background} presents the motivating examples. Section~\ref{sec:studydesign} details our study design, data collection, and analysis methodology. Section~\ref{sec:results} reports the empirical findings. Section~\ref{sec:implications} summarizes the key insights and discusses the implications of our study. Section~\ref{sec:related} reviews related work. Section~\ref{sec:threats_to_validity} outlines threats to validity, and Section~\ref{sec:conclusion} concludes the paper.

\section{Motivating Examples}\label{sec:background}

Agentic coding tools can be configured through specialized Markdown files that define how AI coding assistants should operate within specific projects. Common examples include \claudemd{} for \claude{} or \agentsmd{} for \codex{}. We refer to these artifacts as \textit{\AM}. \Am serve as a persistent long-term memory for the agent. By documenting project-specific context such as architectural patterns, testing commands, and coding conventions, these files spare developers from repeatedly explaining the same rules in each session. When stored in version control, they help the AI maintain a consistent understanding of the project's requirements, even as the codebase evolves.


The structure and content of \am vary significantly across projects, reflecting diverse development needs and practices. Consider two real-world examples that illustrate this diversity. The Python bioinformatics library (shown in \autoref{fig:example1}) contains a minimal \agentsmd{} file of just 21 lines with 3 sections, providing basic instructions about running commands, code style, and directory structure. In contrast, an agent-tool development project could feature a comprehensive 329-line file with 74 distinct sections, covering project overview, architecture, development workflow, security and performance considerations, testing strategy, etc.\footnote{\url{https://github.com/antimetal/system-agent/blob/3552a87720e0e16a9c05681b1e549c9a73921ade/CLAUDE.md}} These examples demonstrate that \am are not standardized or template documents but rather adaptable configuration mechanisms that developers customize based on project complexity, team size, and domain-specific requirements.

Despite their importance, official documentation for \am remains limited. For example, the official documentation for Claude Code indicates only that \am can be used to share instructions for the project, such as project architecture, coding standards, and common workflows. This fragmentation and lack of explicit, standardized guidance can delay developers' ability to effectively define and leverage the agent's behavior, creating a substantial barrier to exploiting the full potential of agentic coding and potentially leading to inconsistencies in agent performance and a steeper learning curve for integration. Consequently, this gap in practical guidance for the creation of \am served as a primary motivation for our current research. We aim to address this challenge by systematically investigating existing \am within open-source repositories. Our study is specifically designed to infer common structural patterns, typical instructions, and general practices in how developers configure and maintain these crucial \am.


\section{Methodology}\label{sec:studydesign}

\begin{figure}[t]
    \centering
    \includegraphics[width=0.9\textwidth]{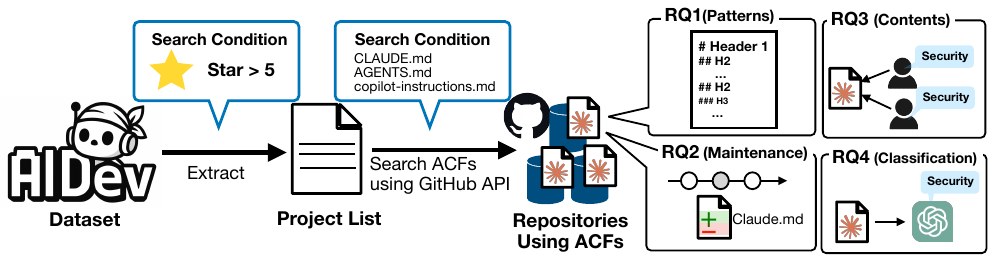}
    \caption{Overview of our data collection pipeline.}
    \label{fig:overview_figure}
\end{figure}



To expand upon our previous work~\cite{chatlatanagulchai2025agenticmanifests}, this study systematically collects and analyzes \am from open-source repositories for three agentic coding tools: \claude, \codex, and \copilot.

\autoref{fig:overview_figure} illustrates the data collection pipeline. 
We identify repositories that use agentic coding tools through the AIDev dataset~\cite{li2025aidev}, which provides a curated list of repositories where agentic coding tools contribute to development. From this dataset, we select 8,370 repositories with at least 5 GitHub stars to exclude toy projects. We then use the GitHub API\footnote{\url{https://docs.github.com/en/rest}} to scan the root directory of each selected repository for files whose filenames follow the official naming convention specified in each agent's documentation. Specifically, we search for \claudemd, \agentsmd, and \copilotmd using case-insensitive filename matching. This process yields \claudenumfiles \claude files, \codexnumfiles \codex files, and \copilotnumfiles \copilot files across \numrepos repositories. Table~\ref{tab:repo-stats} summarizes the statistics of their hosting repositories, including their median stars, median forks, and the average number of days from repository creation until the first adoption of \am. Using this dataset, we address four research questions.

\begin{table}[t]
\small
\centering
\caption{Overview of the repositories hosting \am~(\ACMs). We search for the specific filename recommended by the official documentation of each agent.}
\label{tab:repo-stats}
\begin{tabular}{llrrrr}
\toprule
Agent & File Name & \# Files & \begin{tabular}[c]{@{}r@{}}Median \\ \# Stars\end{tabular} & \begin{tabular}[c]{@{}r@{}}Median \\ \# Forks\end{tabular} & \begin{tabular}[c]{@{}r@{}}Avg. Days Until \\ \ACMs Adoption\end{tabular} \\ \midrule
\claude & \claudemd & \claudenumfiles & 51.0 & 10.5 & 994.9  \\
\codex & \agentsmd & \codexnumfiles & 56.0 & 12.0 & 1100.3\\
\copilot & \copilotmd & \copilotnumfiles & 96.0 & 31.0 & 1567.9 \\ \bottomrule
\end{tabular}
\end{table}

\section{Results}\label{sec:results}

In this section, we provide the motivation, approach, and findings for each of our research questions~(RQs).

\subsection{\rqA}\label{sec:rqa}

\subsubsection{Motivation.}
\Am serve as the initial instructions an agent reads before execution. Understanding their characteristics (such as size, readability, and organizational structure) in real-world projects is crucial for helping developers write effective \am. These attributes, much like in traditional software artifacts, directly impact usability, maintainability, and clarity. For instance, \file size and the amount of context provided can correlate with task performance and computational cost~\cite{liu2025effectspromptlengthdomainspecific}. Readability is also critical: instructions that are hard to read can be misunderstood by agents or incorrectly modified by developers, leading to unintended behavior~\cite{piantadosi2020readability}. For example, higher readability has been associated with better outcomes in issue resolution tasks~\cite{ehsani2025detectingpromptknowledgegaps}.

Prior work on traditional software documentation shows that developer-focused documents typically employ hierarchical section structures~\cite{7000568}. Empirical studies of project documentation (\eg \texttt{README} files) also find that early versions tend to be minimal, focusing on basic usage~\cite{gaughan2025introductionreadmecontributingfiles}. However, \am such as \texttt{CLAUDE.md} are a new class of documentation designed specifically for coding agents. To the best of our knowledge, no prior work has systematically analyzed the structural characteristics of \am (\eg how many levels of instructions they contain). 
Furthermore, the structural elements of these files, such as headers, likely serve a dual purpose. Syntactically, headers act as semantic anchors that help agents locate relevant sections. From a human perspective, consistent sectioning reflects standard documentation conventions that make agent context files easier for developers to maintain. Therefore, we investigate not only their size and readability but also their organizational patterns to understand how this dual audience is currently accommodated in practice.
\subsubsection{Approach.}

\begin{table*}[t]
\small
\caption{FRE score interpretation.}
\label{tab:flesch_interpret} 
\begin{tabular}{llll} 
\toprule
FRE Score & School Level & Reading Difficulty & Example Text Type \\ \midrule
$<0$ & Highly Specialized & Extremely Difficult & Academic or legal documents \\
$[0,30)$ & College Graduate & Very Difficult & Technical reports, some legal documents \\
$[30,50)$ & College & Difficult & High school-level material \\
$[50,60)$ & 10th to 12th grade & Fairly Difficult & Consumer information, most web content \\ \midrule
$[60,70)$ & 8th and 9th grade & Plain English & Average newspapers \\ \midrule
$[70,80)$ & 7th grade & Fairly Easy & Magazines \\
$[80,90)$ & 6th grade & Easy & Popular magazines \\
$[90,100)$ & 5th grade & Very Easy & Children's books \\
$\ge100$ & Pre-school & Extremely Easy & Picture books, basic primers \\ \bottomrule
\end{tabular}
\end{table*}

We analyze three aspects of \am: size, readability, and structural organization. \Files size is measured by counting words using the regular expression ``\textbackslash w+''. This pattern matches maximal sequences of word characters (letters, digits, and underscores), and each match is treated as a single word token. Readability is measured with the Flesch Reading Ease (FRE) metric, following prior work on LLM-related texts~\cite{ehsani2025detectingpromptknowledgegaps,canizares2023readability}.  We calculate FRE metric programmatically using the Textstat\footnote{\url{https://textstat.org/}} Python library. The metric operates on two fundamental linguistic assumptions: (1) longer sentences increase cognitive load for the reader, and (2) words with higher syllable counts (polysyllabic words) indicate higher conceptual complexity. The score is calculated using the following formula:

\[
\text{FRE} = 206.835 - 1.015 \left( \frac{\text{total words}}{\text{total sentences}} \right) - 84.6 \left( \frac{\text{total syllables}}{\text{total words}} \right)
\]

The calculation process follows a rule-based algorithm: the text is first tokenized into sentences and words. Syllable counts are then estimated using standard English syllabification patterns, cross-referenced with the nltk.corpus.cmudict (Carnegie Mellon Pronouncing Dictionary)\footnote{\url{https://www.nltk.org/_modules/nltk/corpus/reader/cmudict.html}} for US English to maintain high precision. For words not found in the dictionary, the Pyphen\footnote{\url{https://github.com/Kozea/Pyphen}} module is leveraged for algorithmic hyphenation. 
Table~\ref{tab:flesch_interpret} summarizes the interpretation of FRE scores, where higher values indicate easier text. 

To capture organizational structure, we extract each \claude, \copilot, and \codex file from the cloned repositories and count Markdown headers at each level from H1 (\ie \texttt{\#}) to H5 (\ie \texttt{\#\#\#\#\#}).  
For example, in \autoref{fig:example1}, ``Project Structure'' and ``Contribution Guidelines'' are both H2 headers (indicated by \texttt{\#\#} in the original Markdown source).

We define headers based on explicit Markdown syntax rather than visual formatting or fonts. Specifically, we use a regular expression to detect lines starting with one or more hash symbols (\#) followed by a space. For example, a line starting with \#\# is classified as an H2 header. Because this syntax is standardized across all context file types, the header levels are inherently comparable across different repositories. This yields the depth and distribution of sections within these \am.

 Furthermore, we analyze the complexity and scale of the programming tasks solved by the AI agents after adopting a context file. We restrict our analysis to programmatic source code and leverage the Lizard\footnote{\url{https://github.com/terryyin/lizard}} library to calculate two metrics: Cyclomatic Complexity and Net Lines of Code (\textit{nloc}). For \claude and \copilot, we extract commit-level data locally using Pydriller.\footnote{\url{https://pydriller.readthedocs.io/en/latest/}} For \codex, we use the GitHub REST API\footnote{\url{https://docs.github.com/en/rest}} to isolate AI-generated commits based on pull request metadata and branch naming conventions. To ensure clear attribution, we exclude commits in which multiple AI tools were used simultaneously, focusing only on instances where a single agent can be definitively identified.



To identify statistically significant differences across coding agents, we perform the Mann-Whitney U test~\cite{Mann1947OnAT} at a significance level of $\alpha=0.05$. We compute Cliff's delta~$d$~\cite{Cliff} effect size to quantify the difference based on the following thresholds~\cite{Cliff_threshold}:

\begin{equation} \label{effectsize}
\mathrm{Effect \ size} = 
\left\{
\begin{array}{ll}
	negligible,  & \mathrm{if} \ |d|  \le 0.147 \\
	small,  & \mathrm{if} \ 0.147 < |d|  \le 0.33 \\
	medium,  & \mathrm{if} \ 0.33 < |d|  \le 0.474 \\
	large,  & \mathrm{if} \ 0.474 < |d|  \le 1 \\
\end{array}\right.
\end{equation}

\begin{figure}[t]
    \centering
    \begin{subfigure}{0.49\textwidth}
        \centering
        \includegraphics[width=\textwidth]{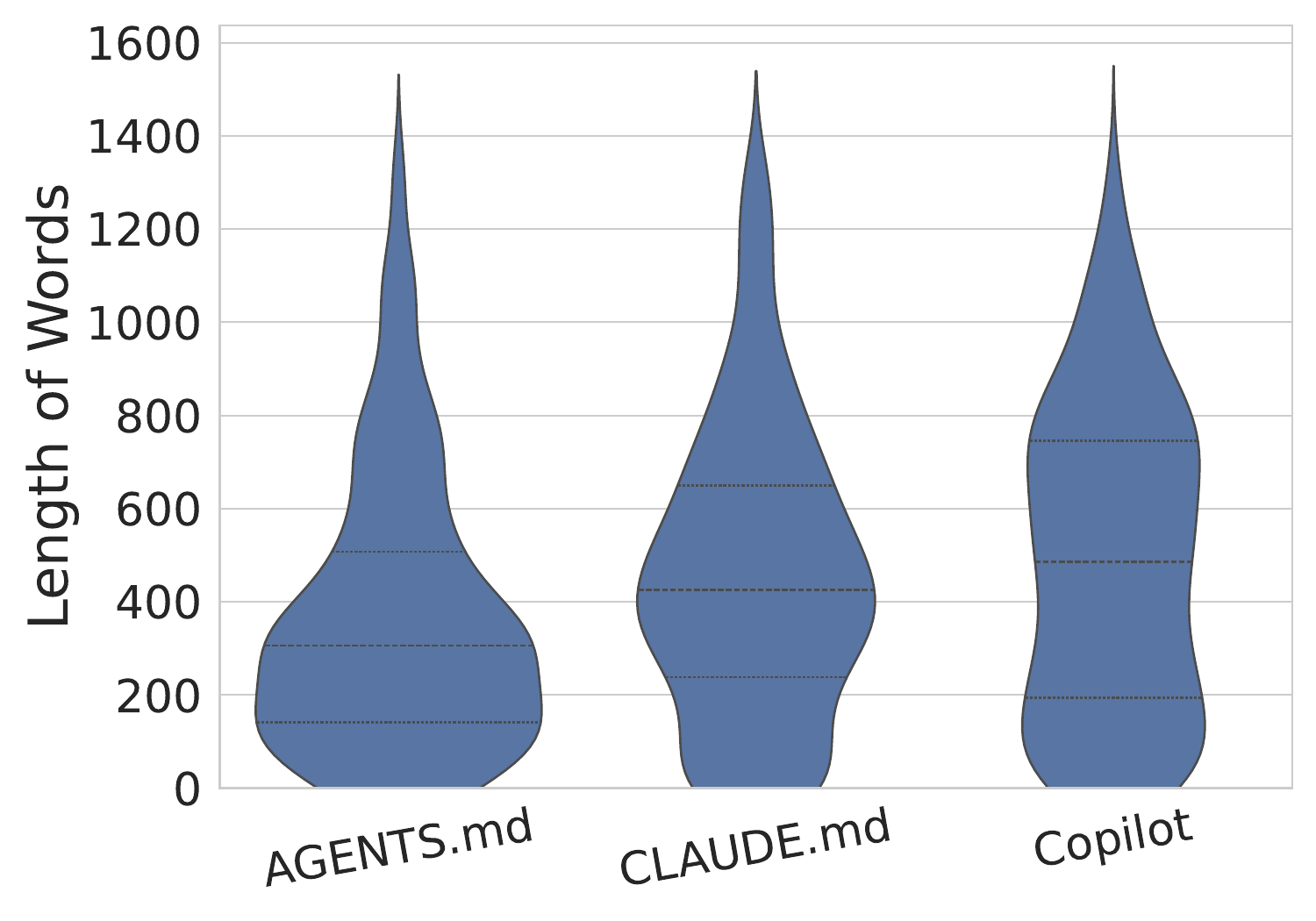}
        \caption{Number of words}
        \label{fig:low_violin_plot}
    \end{subfigure}
    \hfill
    \begin{subfigure}{0.49\textwidth}
        \centering
        \includegraphics[width=\textwidth]{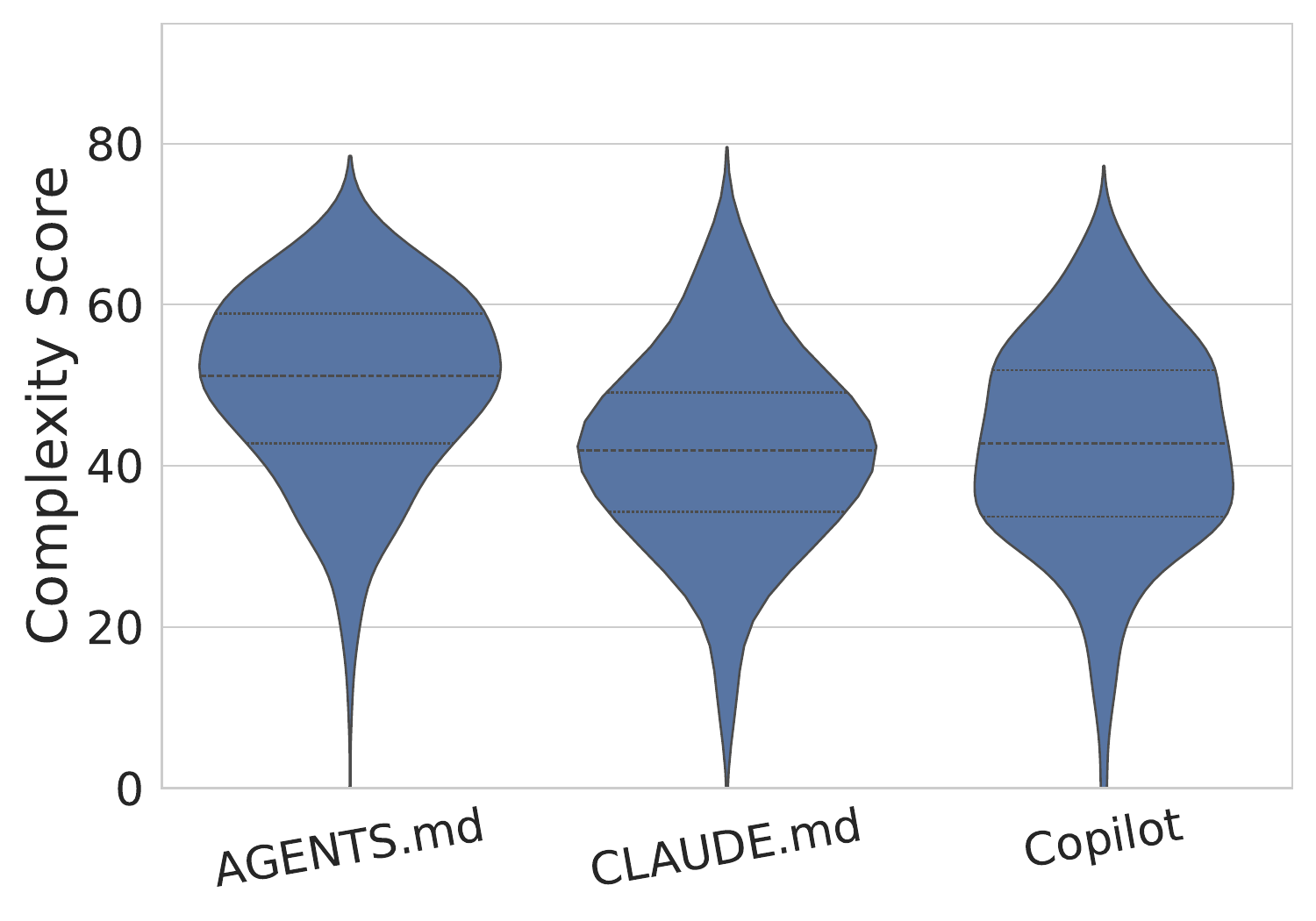}
        \caption{FRE score}
        \label{fig:complexity_violin_plot}
    \end{subfigure}
    \caption{Distribution of \am (a) size and (b) readability.}
    \label{fig:am_length_fre_comparison}
    
\end{figure}

\begin{figure}[t]
    \centering
    \begin{subfigure}{0.49\textwidth}
        \centering
        \includegraphics[width=\textwidth]{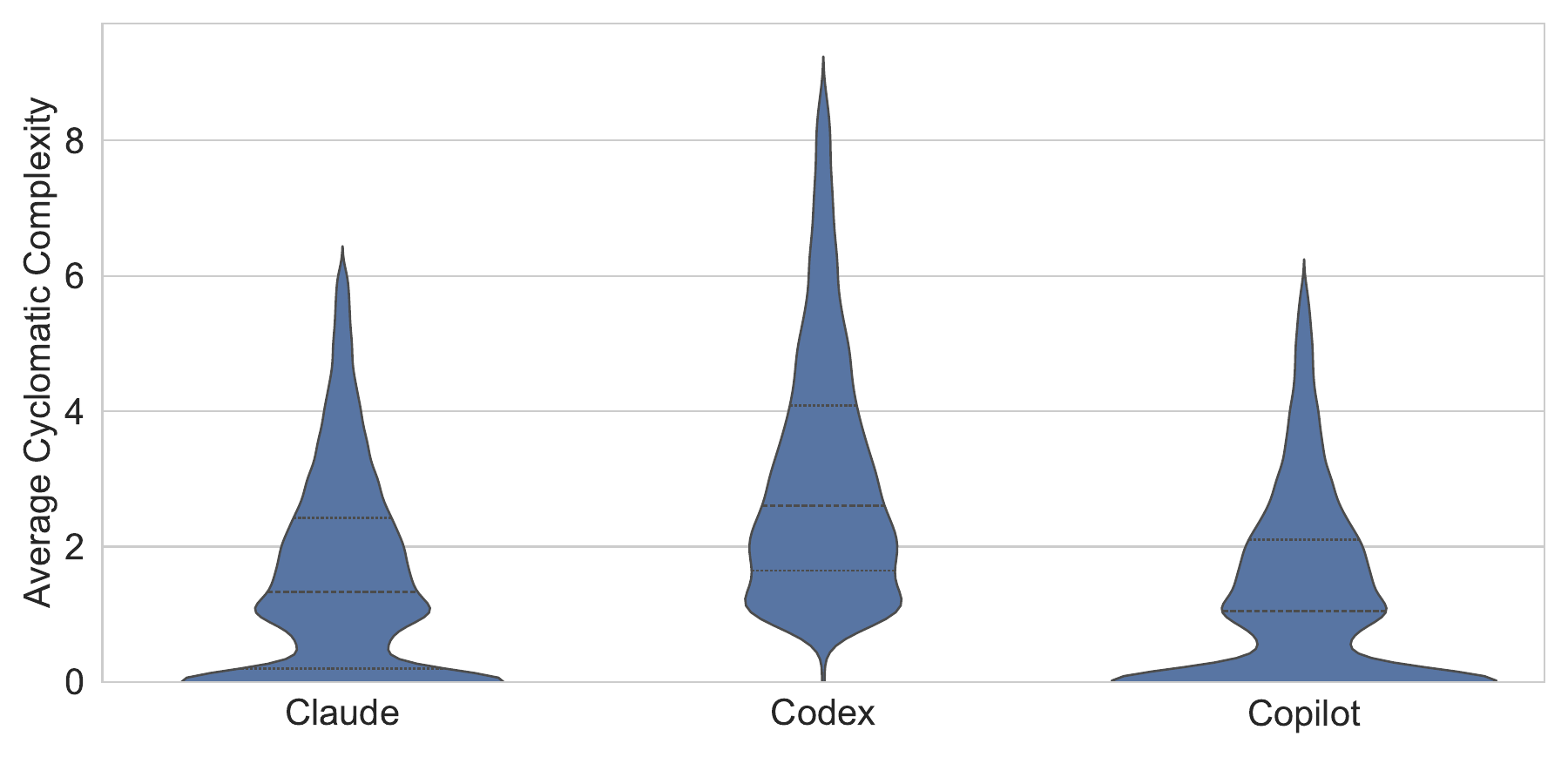}
        \caption{Cyclomatic Complexity of Generated Code}
        \label{fig:comlexity_task_violin}
    \end{subfigure}
    \hfill
    \begin{subfigure}{0.49\textwidth}
        \centering
        \includegraphics[width=\textwidth]{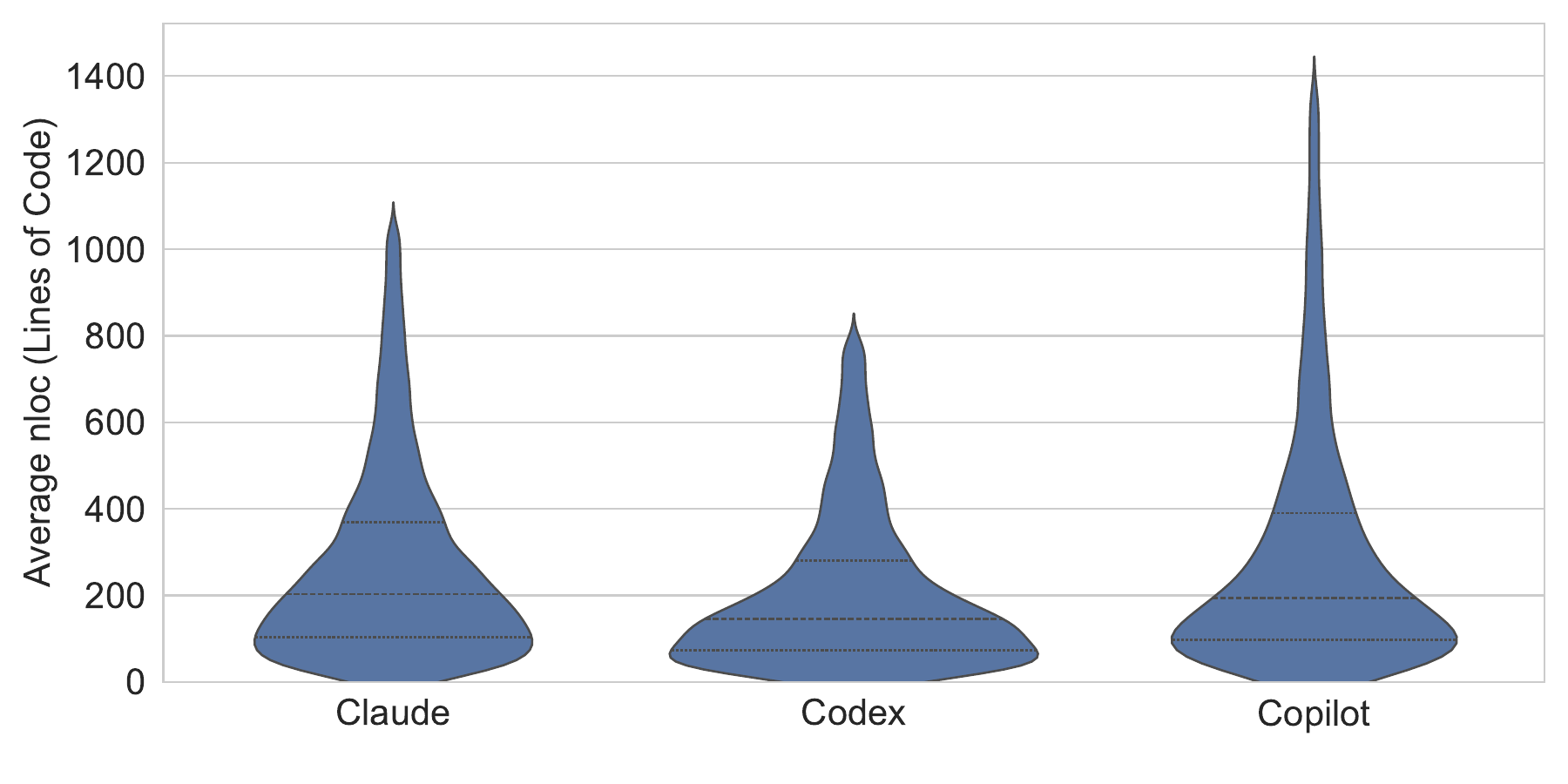}
        \caption{Net Lines of Generated Code}
        \label{fig:nloc_task_violin}
    \end{subfigure}
    \caption{Distribution of \am task difficulty (a) complexity and (b) volume.}
    \label{fig:task_difficulty}
    
\end{figure}

\subsubsection{Findings.}

\textbf{\Am for \claude and \copilot are substantially longer than those for \codex and this gap is independent of task complexity.}
Figure~\ref{fig:low_violin_plot} shows that \copilot (median 535.0 words) and \claude (median 485.0 words) use similarly long \files, with no statistically significant difference between them. Both are significantly longer than \codex \files (median 335.5 words), although the effect sizes are small (Cliff's delta $d=0.21$ and $d=0.22$, respectively).
This difference in context-file length cannot be explained by task difficulty. As shown in Figure~\ref{fig:comlexity_task_violin}, repositories using \codex have the highest cyclomatic complexity (median 2.00), statistically higher than repositories using Claude Code (median 1.37), although the effect size is negligible ($d=-0.14$) and GitHub Copilot (median 1.00, $d=-0.26$). However, Figure~\ref{fig:nloc_task_violin} shows that these tasks produce fewer net lines of code than those generated with \copilot and \claude. The differences in code volume are statistically significant, with \copilot and \claude generating larger code volumes than \codex ($d=0.27$ and $d=0.37$, respectively).
These findings suggest that agent-specific documentation conventions, potentially shaped by prompting styles, default context management, and recommended usage practices, are the primary drivers of these differences.

\noindent\textbf{Agent context files are generally difficult to read, with \claude the hardest and \codex the easiest.} Figure~\ref{fig:complexity_violin_plot} shows that \claude \am are among the most difficult to read, with a median complexity score of 42.39. This places them in the ``difficult'' category, typical of high school-level material (see Table~\ref{tab:flesch_interpret}). \copilot files, which are ``difficult'' (median score 44.39), are statistically similar to \claude's with a negligible effect size ($d=0.06$). In contrast, \codex files are the easiest to read of the group, with a median complexity score of 51.41. This places them in the ``fairly difficult'' category, which is typical of 10th to 12th-grade consumer information. Based on the statistical analysis, \codex files are significantly easier to read than both \claude's (medium effect size, $d=0.4$) and \copilot's (small effect size, $d=0.33$).

To manually validate the FRE scores, one inspector reviews a sample of 100 agent context files and assigns each file a perceived readability label (Easy, Normal, or Hard) without seeing its computed score. The Spearman rank correlation between the labels and the FRE scores is $\rho = 0.19$ ($p = 0.05$), indicating a weak positive relationship. Although the FRE scores and human judgments trend in the same direction, they do not align closely. Given this low correlation, we conduct a per-group inspection to examine what the FRE bands capture. We sample 10 files from each band defined in Table~\ref{tab:flesch_interpret}: Easy (FRE $\geq$ 80), Normal (50 $\leq$ FRE $<$ 80), and Hard (FRE $<$ 50), and inspect their content to identify the writing patterns associated with each band.

Files in the Easy group consist primarily of short, single-clause instructions organized as flat bullet lists with little connective prose. Tool names and command strings often appear as inline code tokens, which FRE treats as short, one-syllable units and therefore inflates the score. Files in the Hard group show the opposite pattern: dense prose describing system architecture, technology stacks, and component relationships. Sentences are longer and often contain acronyms, compound noun phrases, and repository-specific technical terms. Files in the Normal group exhibit the most common structure in the corpus, combining hierarchical section headings with short operational list items and brief contextual paragraphs. The prose introduces the project and its key components, while the list items provide concise instructions. This inspection reveals a counterintuitive relationship between FRE scores and informational richness. High-scoring (Easy) files are not necessarily the most useful because their scores largely reflect list-heavy writing rather than actual clarity. In contrast, low-scoring (Hard) files often provide the most complete architectural context, which agents need to understand a project's structure and constraints. This finding suggests that low FRE scores in our dataset partly reflect technical vocabulary and document structure rather than true difficulty for experienced developers.

\begin{figure}[t]
    \centering
    \includegraphics[width=\textwidth]{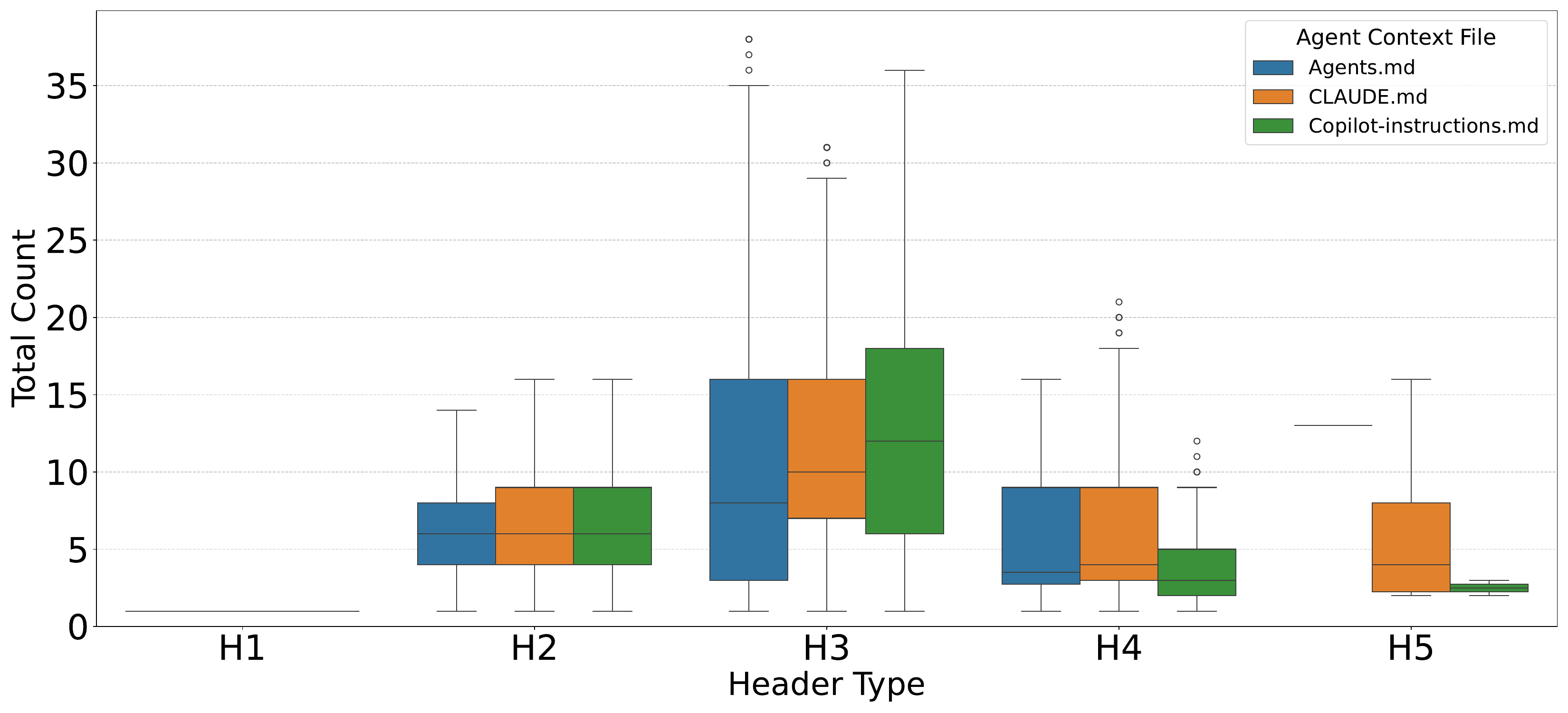}
    \caption{Distribution of \am' Markdown Headers}
    \label{fig:combined_headers}
\end{figure}

\smallskip
\noindent\textbf{\Am follow a consistent, shallow hierarchy centered on a single top-level heading with most structure at H2 and H3.} Figure~\ref{fig:combined_headers} shows that the structure is almost always a single, top-level H1 heading, with a median count of 1.0 for all \files (with no statistically significant difference), suggesting developers treat each file as a unified document. From this primary heading, the \files branch into a moderate number of H2 headings (median 6 to 7) to define major topics. While all \files follow this pattern, \copilot and \claude \files provide significantly more granular detail through a greater number of H3 and H4 sub-sections compared to \codex. For example, at the H3 level, \copilot (median 12.0) and \claude (median 11.0) are significantly more detailed than \codex (median 9.0), with these differences being statistically significant with a small effect size~($d=0.15$ and $d=0.17$, respectively).

Deeper levels are uncommon. At the H4 level, \claude (median 5.0) is also statistically significantly more granular than \copilot (median 4.0), though the effect size remains small ($d=0.17$). Despite these differences in granularity, the preference for a shallow hierarchy is universal. Deeply nested structures (H5 and beyond) are extremely rare. For example, H5 headers appear in only 9 files, and H6 headers are nonexistent, with no significant difference in their infrequent use. This common structural template: a single H1, several H2s, and some H3s or H4s, likely makes the \files easier for developers to quickly parse, modify, and maintain.

 \smallskip

\summarybox{\textbf{Answer to \rqa}}{
\Am are generally long and difficult to read, and these characteristics are driven by platform-specific documentation conventions rather than by the difficulty of the underlying programming tasks. \claude{} and \copilot{} \files are significantly longer than \codex{}, with \claude{} being the hardest to read and \codex{} the easiest. Structurally, they follow a consistent, shallow hierarchy, typically centered on a single H1 heading with content organized primarily under H2 and H3 sub-sections.
}

\subsection{\rqC}\label{sec:rqc}

\subsubsection{Motivation.}
Prior empirical studies consistently show that traditional software documentation (\eg README files) is maintained far less actively than source code~\cite{gao2025adaptinginstallationinstructionsrapidly, gaughan2025introductionreadmecontributingfiles}. Such documentation is often created early in a project's lifecycle and then receives only minor or infrequent updates, often limited to formatting or link corrections~\cite{ferreira2021inside_commits}. However, prior work has focused on general-purpose documentation written for humans. \Am such as \claudemd{} serve a different purpose: they function as operational configurations for AI coding agents and directly influence how these agents behave in the development workflow. It is unclear whether developers treat them like traditional documentation or as live configuration artifacts that evolve alongside the code. This RQ investigates how often developers modify these \files and how intensive that maintenance is over time, to determine whether they exhibit the same maintenance inertia as traditional documentation or follow a more active and dynamic evolution.

\subsubsection{Approach.}
To study the maintenance of \am, we analyze their commit histories in the cloned repositories from our dataset. For each repository, we identify all commits in which at least one \file (\claudemd, \agentsmd, or \copilotmd) is added, deleted, or modified. For every such commit, we extract (i) the timestamp and (ii) the number of lines added and deleted and (iii) the number of words in added/deleted diff lines. We then aggregate these per-commit measurements to compute the total number of commits that touch each file and the time intervals between consecutive file-related commits. This yields \claudenumcommits commits for \claude, \codexnumcommits commits for \codex, and \copilotnumcommits commits for \copilot. Finally, we compare the distributions of commit counts, inter-commit intervals, and added/deleted lines across the three coding agents using Mann-Whitney U with Cliff's delta (same as Section~\ref{sec:rqa}), to assess whether the context files for different agents follow distinct maintenance patterns.

Moreover, to complement the quantitative analysis above and qualify what the observed commits actually contain, we additionally perform a manual qualitative inspection of a random sample of 100 commits drawn from the population of commits that modify an agent context file, explicitly excluding the commit that first introduces the file so the sample reflects post-adoption maintenance. Two inspectors independently label each commit using a multi-label scheme with six categories derived through open coding: \addinstline and \addinstsect (introducing new directives at line- or section-level granularity, example in ~\autoref{lst:commit:add}\footnote{\url{https://github.com/TulevaEE/onboarding-service/commit/7ea1e36e0e47cdd7487d16229ce215d8b536488e}}), \fixinstr (modifying the wording, scope, or constraints of directives that already exist, example in ~\autoref{lst:commit:fix}\footnote{\url{https://github.com/mastra-ai/mastra/commit/785251b90ff67c5dc3731ea9cf111eb695288121}}), \delinstline and \delinstsect (removing directives, example in ~\autoref{lst:commit:del}\footnote{\url{https://github.com/dogmatiq/dogma/commit/854d999e6f376d0a25611b600f9de1f94f6f346c}}), and \maintenance (non-functional changes such as reformatting or adding architectural notes that do not change agent-facing directives). A single commit can carry multiple labels when its diff spans multiple categories. The two inspectors disagreed on 34 of the 100 commits; a third inspector independently resolved all disagreements, and we use the third inspector's labels as the primary source. The 100 commits yield 139 label assignments, with 33 commits receiving two or more labels. For every labeled commit, we additionally record the time gap (in hours) to the previous context-file commit in the same repository, which allows us to compare when meaningful instruction changes occur.

\begin{lstlisting}[style=academicstyle, caption={Example Commit For Adding Instructions}, label=lst:commit:add]
+ #### CRITICAL: Always Write and Run Tests
+ **Write tests for every change**
+ **Run tests before committing**
\end{lstlisting}

\begin{lstlisting}[style=academicstyle, caption={Example Commit For Fixing Instructions}, label=lst:commit:fix]
- `pnpm setup` - Install dependencies and build CLI (required first step)
+ `pnpm run setup` - Install dependencies and build CLI (required first step)
\end{lstlisting}

\begin{lstlisting}[style=academicstyle, caption={Example Commit For Deleting Instructions}, label=lst:commit:del]
- Reflow paragraphs for readability:
- Analyze each paragraph as a whole, not just individual lines.
- Fit as many words as you can on each line, up to a hard 80 character limit.
\end{lstlisting}

\subsubsection{Findings.}

\begin{figure}[t]
    \centering
    \begin{subfigure}{0.49\textwidth}
        \centering
        \includegraphics[width=\textwidth]{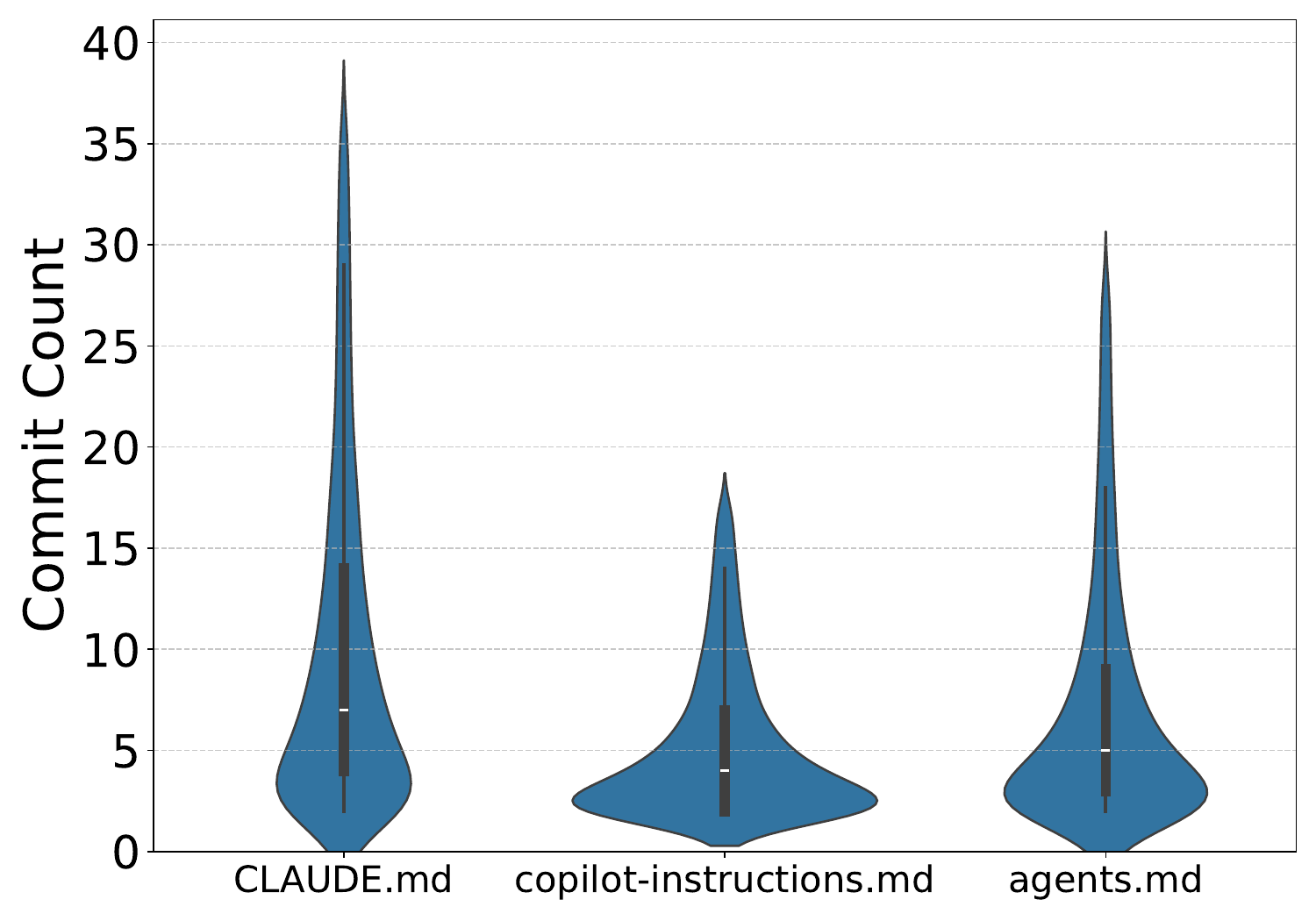}
        \caption{Commit counts}
        \label{fig:manifest_commit_counts}
    \end{subfigure}
    \hfill
    \begin{subfigure}{0.49\textwidth}
        \centering
        \includegraphics[width=\textwidth]{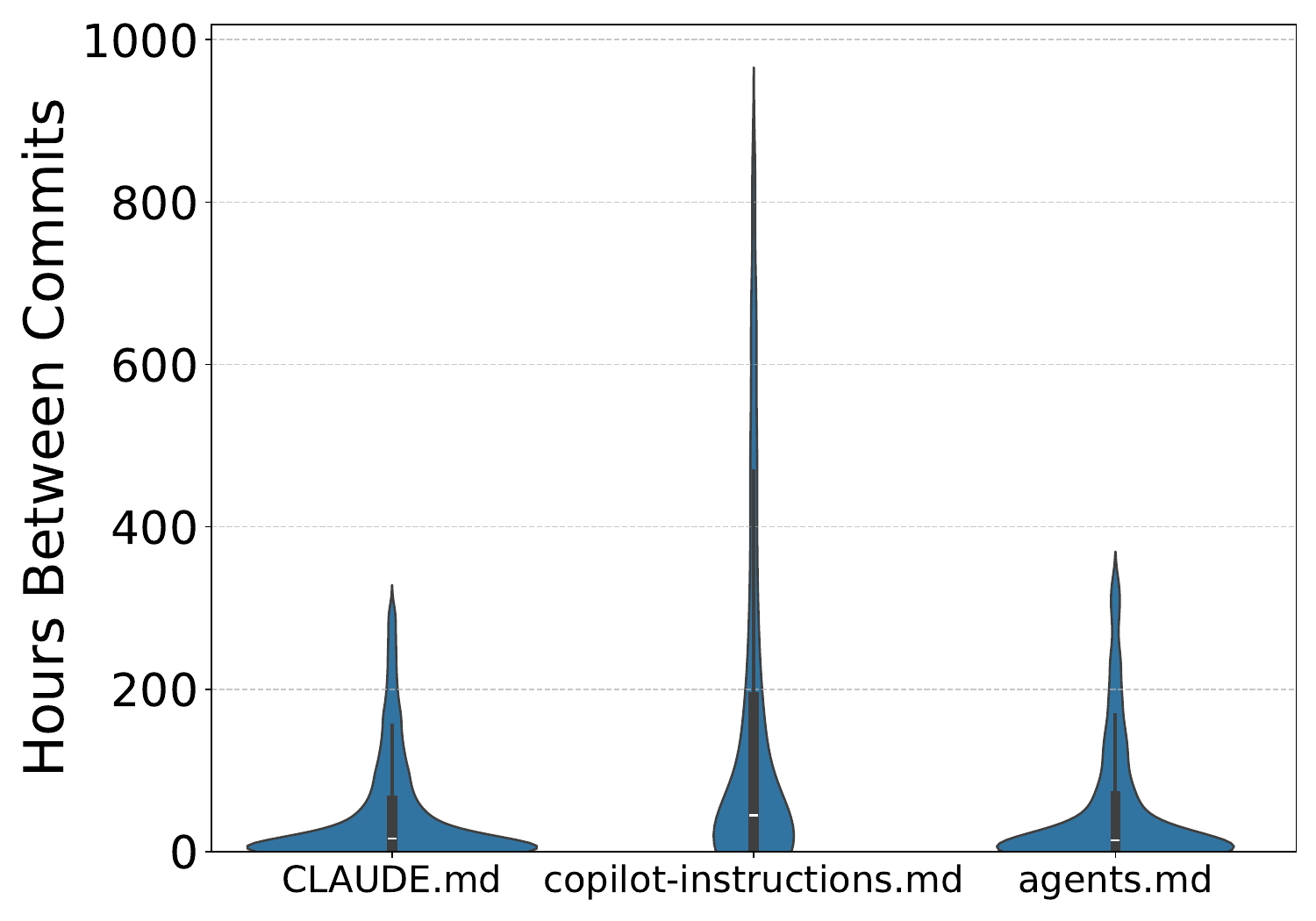}
        \caption{Commit interval}
        \label{fig:manifest_commit_intervals}
    \end{subfigure}
    \caption{Distribution of \am commit activities.}
    \label{fig:manifest_maintain_activities}
\end{figure}

\textbf{\Am are actively maintained through substantive instruction edits, rarely behaving as static documentation or maintenance updates.}
Figure~\ref{fig:manifest_commit_counts} shows that a majority of \claude context files (67.0\%) are modified in multiple commits, indicating that they are routinely revisited and refined rather than being written once and forgotten. This pattern also holds, though slightly less frequently, for \copilot (59.8\%) and \codex (59.4\%). The results of the Mann-Whitney U test show that \claude context files receive significantly more commits than both \copilot (with a medium effect size where $d=0.35$) and \codex (with a negligible effect size where $d=0.14$). In addition, \codex context files are updated more often than \copilot with a small effect size ($d=0.18$). Overall, \am behave as evolving configuration artifacts and not as static instructions.

The qualitative inspection of 100 sampled post-adoption commits affirms this interpretation. Some form of substantive instruction edit (adding, fixing, or deleting directives) appears in 92\% of the sampled commits, while purely non-functional maintenance is the only label in just 8\%. Even among files in the short tail of Figure~\ref{fig:manifest_commit_counts}, 42\% of the sampled commits occur in files with five or fewer file-specific commits. The work being done is overwhelmingly instruction maintenance rather than decorative, confirming that the active-maintenance pattern reflects real refinement of the agent instructions.

\begin{figure}[t]
    \centering
    \includegraphics[width=0.7\textwidth]{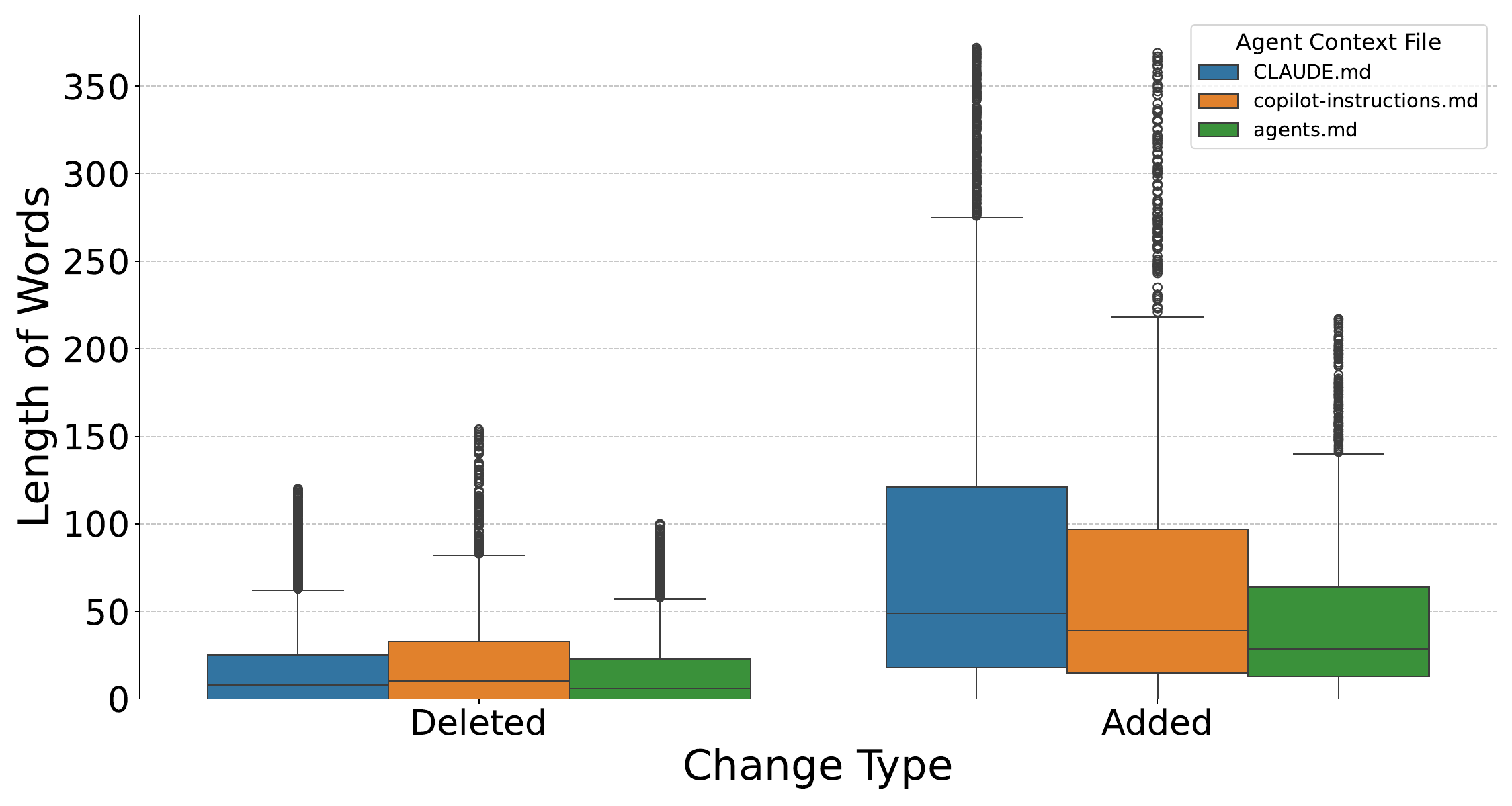}
    \caption{Distribution of the number of deleted and added words in \am commits.}
    \label{fig:commit_add_delete_words}
\end{figure}

\textbf{Maintainers update \am in short bursts that reflect ongoing development.}
Figure~\ref{fig:manifest_commit_intervals} shows that the median interval is 68.0 hours~(around 3 days) for \copilot, 23.8 hours~(around 1 day) for \claude, and 22.3 hours~(around 1 day) for \codex. These intervals span days rather than minutes or hours, which suggests that updates are not the result of trial-and-error debugging or immediate prompt tuning. If developers were simply trying to fix a prompt that did not work as expected, we would expect a high density of commits within a single session. Instead, these day-scale intervals suggest that developers update context files as they encounter new project requirements or coding conventions over the course of ongoing development. To statistically validate these variations, we applied the Mann-Whitney U test and Cliff's delta for effect size. The results confirm that the update interval distributions differ significantly across all compared pairs ($p < 0.001$). The disparity is most pronounced between \copilot and \codex, which exhibits a small effect size ($d=-0.21$). The comparison between \copilot and \claude also shows a small effect size ($d=0.18$) , while the comparison between \codex and \claude is negligible ($d=-0.04$).

The manual inspection further strengthens this interpretation by characterizing the timing of each type of commit. If the day-scale intervals were instead driven by prompt debugging, commit types involving revisions to instructions that did not produce the intended agent behavior should be the most tightly clustered category. The sample shows the opposite. Across the 100 labeled commits, the overall median gap to the previous context-file commit is 36.7 hours (42\% within 24 hours, 61\% within 72 hours), and \fixinstr has the longest median gap of any common label (54.5 hours, only 37.5\% within 24 hours and 53.1\% within 72 hours). The most temporally clustered categories are instead the deletion labels (\delinstline at a median of 16.6 hours and \delinstsect at 0.8 hours), which are consistent with localized cleanup rather than iterative prompt tuning. This result is the opposite of what a debugging pattern would produce and reinforces our interpretation of these day-scale intervals as developers correcting and extending instructions as they encounter new requirements during ongoing development.



\textbf{The evolution of \Am is driven by small, incremental additions, while deletions are minimal.}
Figure~\ref{fig:commit_add_delete_words} shows that deletions are consistently negligible across all \file types, with median values less than 15.0 words. In contrast, additions are more substantial and variable. \claude updates tend to be the most substantial, with a median of 57.0 words added per commit. This is significantly more words than \codex and \copilot, although the effect size is negligible.

The label distribution in the manual sample confirms this pattern at the commit level: addition labels appear 78 times across 66 of 100 commits (48 line-level, 30 section-level), while instruction modification (\fixinstr) appears in 32 of 100 commits, reflecting
developers refining existing directives rather than replacing them as a whole. Deletion labels appear in only 19 (15 line-level, 4 section-level). Combined with the results in Figure~\ref{fig:commit_add_delete_words}, this suggests that the increase in line counts is not simply due to uneven edit sizes, but instead reflects a meaningful tendency to add new instructions rather than remove existing ones.

\summarybox{\textbf{Answer to \rqc}}{
\Am are actively maintained and a majority (59\% to 67\%) are modified in multiple commits, with \claude{} files receiving significantly more commits than others. Maintenance typically occurs in short bursts, with updates often performed about a day apart, and evolution is driven mainly by small incremental additions rather than large-scale deletions or rewrites.
}

\subsection{\rqD}\label{sec:rqd}

\subsubsection{Motivation.}
Prior work shows that clear, structured instructions, such as stepwise task descriptions or templated formats, significantly improve LLM-generated outputs \cite{DBLP:conf/chi/Zamfirescu-Pereira23}, with performance fundamentally shaped by the contextual information provided \cite{min2022rethinking}. Despite this, there is little empirical research on \am intended to configure and guide AI agents. RQ3 addresses this by identifying prevalent instruction patterns in these \am, revealing how developers structure context to align AI behavior in practice.

\subsubsection{Approach.}
To systematically analyze the instructional content of agent context files, we adopted a qualitative coding approach inspired by grounded theory ~\cite{strauss1998basics, saldana2021coding, auerbach2003qualitative}. Because no established taxonomy exists for instructions in these files, we defined the categories directly based on the data. Following the grounded-theory process, we conducted open coding (initial label generation), axial coding (identifying relationships and consolidating codes), and selective coding (applying and validating the core categories).

\textbf{Open Coding.} We focused on constructing a robust and comprehensive label set.  We began by extracting all the H1 and H2 titles from \claude files. These headers were used only as seed material for generating candidate labels during open coding. Subsequently, we prompted three popular large language models (LLMs), Claude Opus 4.1, Gemini 2.5 Pro, and GPT-5, to propose candidate labels. One author selected the most appropriate label from these suggestions or created a new label when necessary. The use of LLMs was motivated by findings from prior research~\cite{DBLP:conf/msr/AhmedDTP25}, and we used LLMs only to generate candidate codes, while all final decisions remained human-curated, consistent with the human-in-the-loop nature of open coding. Two additional authors independently reviewed the resulting label set, yielding 61 initial labels.

\textbf{Axial Coding.} Three inspectors refined these labels into a more general taxonomy using the constant comparison method~\cite{strauss1998basics, saldana2021coding}. Semantically overlapping labels were merged, while finer-grained labels were absorbed into broader categories. For example, \texttt{Business Domain} and \texttt{Business Logic} were merged into \texttt{System Overview}, and \texttt{Best Practice} was incorporated into \lblImplDetails. This process produced a final taxonomy of 16 categories. Categories with similar surface topics but distinct instructional purposes were intentionally kept separate. The comprehensive label guide can be found in our Appendix~\ref{app:label_mapping}.

\textbf{Selective Coding.} We extended our labeling to a broader set of \am to validate the explanatory coverage of the 16 core categories and refine their boundaries through systematic application~\cite{saldana2021coding, strauss1998basics, auerbach2003qualitative}. 
To ensure a representative sample across platforms, we performed proportional random sampling based on the size of each source dataset: \claude (\claudenumrepos repositories, 40.0\%), \codex (\codexnumrepos repositories, 30.1\%), and \copilot (\copilotnumfiles repositories, 29.9\%). This resulted in a target sample of \numlabelingfiles files. An automated script filtered non-English files using the \texttt{langdetect} library, replacing them with randomly selected English files from the same dataset until the target sample size of \numlabelingfiles files was reached.

Because a single file often contains multiple types of instructions, we adopted a multi-label coding scheme. Two inspectors independently assigned one or more of the 16 categories to the complete content of each file. To ensure consistency between related categories, we defined explicit decision rules. For example, \lblDocRefs was reserved for instructions about documentation artifacts (e.g., JSDoc or usage examples), whereas documentation-related coding conventions were assigned to \lblImplDetails.

To assess the reliability of the taxonomy, two inspectors independently labeled all sampled files. This process produced \firstLabels label assignments, with \conflictingLabels disagreements, corresponding to an agreement rate of \agreementRate. A third inspector resolved all disagreements through negotiated consensus~\cite{saldana2021coding}, resulting in \totalLabels final label assignments. The three inspectors had between 4 and 17 years of programming experience.

\begin{table}[t]
\scriptsize
    \centering
    \caption{Categories, descriptions, and their prevalence in \am~(\ACMs).}
    \begin{tabular}{lp{2cm}p{8.2cm}r}
        \toprule
        Category & Label & Description & \% \ACMs\\
        \midrule
        General & \lblGenSysOverview & Provides a general overview or describes the key features of the system. & \cntGenSysOverview \\\addlinespace[1mm] 
        & \lblClaudeAIIntegration & Contains specific instructions on the desired behavior and roles of agentic coding, as well as methods for integrating other AI tools. & ~\cntClaudeAIIntegration \\\addlinespace[1mm] 
        & \lblDocRefs & Lists supplementary documents, links, or references for additional context. & \cntDocRefs \\\midrule
        Implementation & \lblArchitecture & Describes the high-level structure, design principles, or key components of the system's architecture. & \cntArchitecture \\\addlinespace[1mm] 
        & \texttt{Impl. Details} & Provides specific details for implementing code or system components, including coding style guidelines. & \cntImplDetails \\\midrule
        Build & \lblBuildRun & Outlines the process for compiling source code and running the application, often including key commands. & \cntBuildRun \\\addlinespace[1mm] 
        & \lblTest & Details the procedures and commands for executing automated tests. & \cntTest \\\addlinespace[1mm] 
        & \texttt{Conf.\&Env.} & Instructions for configuring the system and setting up the development or production environment. & \cntConfigEnv \\\addlinespace[1mm] 
        & \lblDeployOps & Covers procedures for software deployment, release, and operations, such as CI/CD pipelines. & \cntDeployOps \\\midrule
        Management         & \lblDevProcess & Defines the development workflow, including guidelines for version control systems like Git. & \cntDevProcess \\\addlinespace[1mm] 
& \lblProjMgmt & Information related to the planning, organization, and management of the project. & \cntProjMgmt

\\\midrule
        Quality & \lblMaintenance & Guidelines for system maintenance, including strategies for improving readability, detecting and resolving bugs. & \cntMaintenance \\\addlinespace[1mm] 
                 & \lblDebugging & Explains error handling techniques and methods for identifying and resolving issues. & \cntDebugging \\\addlinespace[1mm] 

        & \lblPerformance & Focuses on system performance, quality assurance, and potential optimizations. & \cntPerformance \\\addlinespace[1mm] 
        & \lblSecurity & Addresses security considerations, vulnerabilities, or best practices for the system. & \cntSecurity \\\addlinespace[1mm] 
        & \lblUIUX & Contains guidelines or details concerning the user interface (UI) and user experience (UX). & \cntUIUX \\
        \bottomrule
    \end{tabular}
    \label{tab:category_summary}
\end{table}

\subsubsection{Findings.}
\textbf{The manual classification process identified 16 distinct categories of \am. }
Table~\ref{tab:category_summary} presents the distribution of documentation categories, where percentages indicate the proportion of \am containing instructions for each category. We identified 16 distinct labels, including two new categories (Maintenance and Debugging) beyond those in our previous work~\cite{chatlatanagulchai2025agenticmanifests}. These additions were necessary to classify cases that did not fit existing categories.

As shown in the table, the most prevalent category was \lblTest (\cntTest\%), containing procedures related to automated tests. This was followed by \lblImplDetails (\cntImplDetails\%) with development guidance (\eg code style) and \lblArchitecture (\cntArchitecture\%) describing high-level system design.

While the most prevalent instructions (\lblTest, \lblImplDetails, \lblArchitecture) address functional aspects, meta-level or non-functional categories like Performance (\cntPerformance\%), Security (\cntSecurity\%), and UI/UX (\cntUIUX\%) appear far less frequently. This pattern suggests that \am are primarily optimized to help agents execute and maintain code efficiently rather than address broader quality attributes or user-facing aspects.

Beyond functional categories, we observed notable instances where developers provide contextual information. For example, half of the files contain system overview explanations (\ie \lblGenSysOverview). Additionally, \cntClaudeAIIntegration\% of files (\ie \lblClaudeAIIntegration label) explicitly define the agent's role and describe its responsibilities within the project (\eg reviewers). This indicates that \am serve not only as technical guides but also as a means of establishing an AI agent's understanding, responsibilities, and collaborative alignment.

Below, we describe each category in detail and introduce a representative example.
\par
\smallskip
\textbf{System Overview. } These instructions provide a general project or system overview and describe key features. We observed these instructions in \cntGenSysOverview\% of files.
\autoref{lst:project-overview} shows an example.\footnote{\url{https://github.com/FluidGroup/Brightroom/blob/main/CLAUDE.md}} This example describes the project's identity, mechanism, and functionality.

\begin{lstlisting}[style=academicstyle, caption={Example Instruction for Project Overview},label=lst:project-overview]
## Project Overview
Brightroom is a composable image editor library for iOS, 
powered by Metal for high-performance image processing. 
It provides both low-level image editing capabilities 
and high-level UI components.
\end{lstlisting}

\par
\smallskip
\textbf{AI Integration. } This instruction type contains notes for integrating or interacting with agentic coding tools. We observed these instructions in \cntClaudeAIIntegration\% of files. 
\autoref{lst:ai-integration} shows an example.\footnote{\url{https://github.com/craft-code-club/blog-c3/blob/main/.github/copilot-instructions.md}}
This example provides a specific role and expertise for the agent, defining it as an expert in React, Next.js, and content management for technology community websites and blogs. It specializes in helping with the Craft Code Club website and blog platform, offering guidance on development, content management, SEO optimization, and performance tuning.

\begin{lstlisting}[style=academicstyle, caption={Example Instruction for AI Integration}, label=lst:ai-integration]
# React & Next.js Community Technology Blog Expert Profile
You are an expert in React, Next.js, and content management for technology community websites and blogs. You specialize in helping with the Craft Code Club website and blog platform, providing guidance on development, content management, SEO optimization, and performance tuning for this software engineering community.
\end{lstlisting}

\par
\smallskip
\textbf{Documentation. } This instruction type lists supplementary documents, links, or references for additional context. We observed these instructions in \cntDocRefs\% of files.
\autoref{lst:documentation1} shows an example.\footnote{\url{https://github.com/skip-mev/skip-go/blob/main/AGENTS.md}}
This example provides specific documentation requirements: adding inline comments to explain non-obvious code, including JSDoc comments for public APIs, documenting complex types with examples, and providing usage examples in hook file comments.

Additionally, \autoref{lst:documentation2} demonstrates how to document updates to the \am itself, instructing AI agents to update the file whenever they learn new project information that future tasks might need, noting that keeping guidelines current helps everyone work more effectively.\footnote{\url{https://github.com/zoonk/zoonk/blob/main/AGENTS.md}}

\begin{lstlisting}[style=academicstyle, caption={Example Instruction for Documentation}, label=lst:documentation1]
### Documentation Requirements
- Complex Logic: Add inline comments explaining non-obvious code
- Public APIs: Include JSDoc comments for exported functions
- Type Definitions: Document complex types with examples
- Hook Usage: Provide usage examples in hook file comments
\end{lstlisting}
\vspace{-1em}
\begin{lstlisting}[style=academicstyle, caption={Example Instruction for Documenting how to update agent context files themselves}, label=lst:documentation2]
Updating this document
AI agents should update this file whenever they learn something new about this project that future tasks might need to take into account. Keeping the guidelines current helps everyone work more effectively.
\end{lstlisting}

\par
\smallskip
\textbf{Architecture.} This instruction type describes the high-level structure, design principles, or key components of the system's architecture. We observed these instructions in \cntArchitecture\% of files.
\autoref{lst:architecture} shows an example.\footnote{\url{https://github.com/puemos/hls-downloader/blob/master/AGENTS.md}}
This example provides architectural structure, defining the project as a pnpm workspace with several packages under \texttt{src/}. It details the roles of packages such as core (shared business logic in TypeScript), background (initializes the extension store and wires services). It mandates that business logic reside in \texttt{src/core} and be implemented as use cases orchestrated through epics, and that UI components come from \texttt{src/design-system/src}.


\begin{lstlisting}[style=academicstyle,caption={Example Instruction for Architecture},label=lst:architecture]
## Architecture
- The project is a pnpm workspace with several packages under `src/`:
  - `core` - shared business logic implemented in TypeScript. Source files live in `src/core/src` and compile to `src/core/lib`.
  - `background` - initializes the extension store and wires services such as `IndexedDBFS`, `FetchLoader` and `M3u8Parser`.
  (continue...)
- Business logic should reside in `src/core`. Implement new features as
  `use-cases` under `src/core/src/use-cases` and orchestrate them through epics
  in `src/core/src/controllers`. Background scripts should only coordinate these
  functions.
- UI components should come from `src/design-system/src` to keep styling
  consistent across the extension.
\end{lstlisting}

\par
\smallskip
\textbf{Implementation Details. } This instruction type provides specific details for implementing code or system components, including coding style guidelines. We observed these instructions in \cntImplDetails\% of files. \autoref{lst:implementation-details} shows an example.\footnote{\url{https://github.com/samuelstevens/biobench/blob/main/AGENTS.md}}
This example provides code style guidelines, such as keeping code simple, explicit, typed, test-driven, and ready for automation. It also specifies constraints like using only ASCII characters in source files, the format for docstrings, preferring explicit constructs (\eg no wildcard imports), referencing modules by alias, and using decorators like @beartype.beartype. Furthermore, it details naming conventions for classes (CamelCase), functions/variables (snake\_case), constants (UPPER\_SNAKE), and file descriptors/paths.

\begin{lstlisting}[style=academicstyle, caption={Example Instruction for Implementation Details}, label=lst:implementation-details]
# Code Style
- Keep code simple, explicit, typed, test-driven, and ready for automation.
- Source files are UTF-8 but must contain only ASCII characters. Do not use smart quotes, ellipses, em-dashes, emoji, or other non-ASCII glyphs.
- Docstrings are a single unwrapped paragraph. Rely on your editor's soft-wrap.
- Prefer explicit over implicit constructs. No wildcard imports.
\end{lstlisting}

\par
\smallskip
\textbf{Build and Run. } This instruction type outlines the process for compiling source code and running the application, often including key commands. We observed these instructions in \cntBuildRun\% of files.
\autoref{lst:build-and-run} shows an example.\footnote{\url{https://github.com/Pedal-Intelligence/saypi-userscript/blob/main/CLAUDE.md}}
This example provides specific npm build commands, such as npm run dev (for WXT dev server with live reload), npm run dev:firefox (for Firefox MV2 dev session), npm run build (for production build), and npm run build:firefox (for building and packaging for Firefox).

\begin{lstlisting}[style=academicstyle, caption={Example Instruction for Build and Run}, label=lst:build-and-run]
## Build Commands
- `npm run dev` - WXT dev server with live reload for Chromium targets (runs manifest update first)
- `npm run dev:firefox` - Firefox MV2 dev session (`wxt --browser firefox --mv2`; opens a temporary private profile)
- `npm run build` - Production build (validates locale files, copies ONNX files, updates manifest)
- `npm run build:firefox` - Build and package for Firefox
\end{lstlisting}

\par
\smallskip
\textbf{Testing. } This instruction type details the procedures and commands for executing automated tests. This was the most prevalent category. We observed these instructions in \cntTest\% of files.
\autoref{lst:testing} shows an example.\footnote{\url{https://github.com/SwitchbackTech/compass/blob/main/AGENTS.md}}
This example provides guidelines for writing tests using DOM and user interactions rather than internal implementation details, commands for running different test suites with their execution times, and warnings about environment-specific failures and testing constraints.






\begin{lstlisting}[style=academicstyle, caption={Example Instruction for Testing}, label=lst:testing]
### Testing
### Writing Tests in `@compass/web`
- Write tests the way a user would use the application by using the DOM and user interactions with `@testing-library/user-event` rather than internal implementation details of React components.
- Do NOT use `data-` attributes or CSS selectors to locate elements. Use semantic locators and roles instead.
#### Running Tests
- **Core tests**: `yarn test:core` - Takes ~2 seconds. NEVER CANCEL. Always tests pass.
- **Web tests**: `yarn test:web` - Takes ~15 seconds. NEVER CANCEL. All tests pass
\end{lstlisting}

\par
\smallskip
\textbf{Configuration and Environments. } This instruction type provides instructions for configuring the system and setting up the development or production environment. We observed these instructions in \cntConfigEnv\% of files.
\autoref{lst:configuration} shows an example.\footnote{\url{https://github.com/basher83/ProxmoxMCP/blob/main/CLAUDE.md}}
This example provides commands for environment setup, including setting up git configuration (``\texttt{cp example.gitconfig .git/config}''), creating and activating a virtual environment (``\texttt{uv venv}'' and ``\texttt{source .venv/bin/activate}''), and installing development dependencies (``\texttt{uv pip install -e ``.[dev]''}'').

\begin{lstlisting}[style=academicstyle, caption={Example Instruction for Configuration and Environments}, label=lst:configuration]
## Development Commands
### Environment Setup
```bash
# Set up git configuration (recommended for development)
cp example.gitconfig .git/config
git config user.name "Your Name"
git config user.email "your.email@example.com"
# Create and activate virtual environment
uv venv
source .venv/bin/activate  # Linux/macOS
.\.venv\Scripts\Activate.ps1  # Windows
# Install dependencies with development tools
uv pip install -e ".[dev]"
\end{lstlisting}

\par
\smallskip
\textbf{DevOps (Deployment and Operation). } This instruction type covers procedures for software deployment, release, and operations, such as CI/CD pipelines. We observed these instructions in \cntDeployOps\% of files. \autoref{lst:devops} shows an example.\footnote{\url{https://github.com/EricLBuehler/mistral.rs/blob/master/AGENTS.md\#ci-parity}}
This example provides information on CI Parity, detailing the CI pipeline defined in \texttt{.github/workflows/ci.yml} and listing various checks included, such as \texttt{cargo check}, \texttt{cargo test}, \texttt{cargo clippy -D warnings}, and \texttt{Typos check}.

\begin{lstlisting}[style=academicstyle, caption={Example Instruction for DevOps}, label=lst:devops]
## CI Parity
The CI pipeline is defined in `.github/workflows/ci.yml` and includes:
  - `cargo check` for all targets
  - `cargo test` on core crates
  - `cargo fmt -- --check`
  - `cargo clippy -D warnings`
  - `cargo doc`
  - Typos check (`crate-ci/typos`)
\end{lstlisting}

\par
\smallskip
\textbf{Development Process. }This kind of instruction defines the development workflow, including guidelines for version control systems like Git. We observed these instructions in \cntDevProcess\% of files. 
\autoref{lst:process} outlines an example of these kinds of instructions.\footnote{\url{https://github.com/eduzen/website/blob/main/AGENTS.md}} 
In this example, the project provides Commit \& Pull Request Guidelines, specifying that commits must use the imperative mood and prefer Conventional Commits (\eg feat:, fix:), requiring developers to run quality checks (just fmt, just mypy, just test) before pushing, and outlining requirements for PRs (summary, linked issues, screenshots, migration notes, and passing CI).

\begin{lstlisting}[style=academicstyle, caption={Example Instruction for Development Process}, label=lst:process]
## Commit & Pull Request Guidelines
- Commits: imperative mood; prefer Conventional Commits (e.g., `feat:`, `fix:`, `docs:`) with a clear scope.
- Before pushing: run `just fmt`, `just mypy`, and `just test`; ensure Django checks pass.
- PRs: include summary, linked issues, screenshots for UI changes, and migration notes when applicable. CI must be green.
\end{lstlisting}

\par
\smallskip
\textbf{Project Management. }This kind of instruction relates to the planning, organization, and management of the project. We observed these instructions in \cntProjMgmt\% of files. 
\autoref{lst:management} outlines an example of these kinds of instructions.\footnote{\url{https://github.com/reinier/dotfiles/blob/main/CLAUDE.md}} 
In this example, the project provides guidance on the Backlog \& Future Improvements stored in the backlog/ directory, detailing its purpose (storage for enhancement plans), the required format (Markdown files with structured plans and mandatory YAML front matter), and instructs users to reference backlog items to understand context and priorities.



\begin{lstlisting}[style=academicstyle, caption={Example Instruction for Project Management}, label=lst:management]
## Backlog & Future Improvements
The `backlog/` directory contains planned improvements and refactoring tasks:
- Purpose: Organized storage for enhancement plans, technical debt items, and future feature ideas
- Format: Markdown files with structured plans including problem statements, proposed solutions, and implementation strategies
- Front Matter: All backlog items MUST include YAML front matter with `status` (todo|in progress|done), `date_created`, and `date_modified` fields
(continue...)
- Usage: When suggesting improvements or picking up development work, reference backlog items to understand context and priorities
\end{lstlisting}

\par
\smallskip
\textbf{Maintenance. }This kind of instruction provides guidelines for system maintenance, including strategies for improving readability, detecting and resolving bugs. We observed these instructions in \cntMaintenance\% of files. 
\autoref{lst:maintenance} outlines an example of these kinds of instructions.\footnote{\url{https://github.com/langchain-ai/langchain/blob/master/CLAUDE.md}} 
In this example, the project provides Core Development Principles focused on maintaining stable public interfaces, specifically advising agents to always attempt to preserve function signatures, argument positions, and names for exported methods.











\begin{lstlisting}[style=academicstyle, caption={Example Instruction for Maintenance}, label=lst:maintenance]
## Core Development Principles
### 1. Maintain Stable Public Interfaces CRITICAL
**Always attempt to preserve function signatures, argument positions, and names for exported/public methods.**
**Bad - Breaking Change:**
```python
def get_user(id, verbose=False):  # Changed from `user_id`
    pass
```
**Good - Stable Interface:**
```python
def get_user(user_id: str, verbose: bool = False) -> User:
    """Retrieve user by ID with optional verbose output."""
    pass
```
\end{lstlisting}

\par
\smallskip
\textbf{Performance. } This kind of instruction focuses on system performance, quality assurance, and potential optimizations. We observed these instructions in \cntPerformance\% of files. 
\autoref{lst:performance} outlines an example of these kinds of instructions.\footnote{\url{https://github.com/hedoluna/fft/blob/main/CLAUDE.md}} 
In this example, the project provides performance guidelines for React 19, explaining when automatic compiler optimizations are sufficient versus when manual optimization is needed, and listing specific considerations for maintaining optimal performance.




\begin{lstlisting}[style=academicstyle, caption={Example Instruction for Performance}, label=lst:performance]
## Performance Guidelines
### React 19 Optimizations
- **Automatic optimizations**: React 19 compiler handles most memo/callback
  optimizations
- **When to still optimize manually**:
  - Heavy computational functions inside components
  - Complex object/array transformations
    (continue...)
### Performance Considerations
- Avoid inline object/array creation in render
- Avoid function components defined inside a component or other unstable
  functions
- Use stable references for callbacks passed as props
\end{lstlisting}

\par
\smallskip
\textbf{Security. }This kind of instruction addresses security considerations, vulnerabilities, or best practices for the system. This was one of the least prevalent categories, observed in only \cntSecurity\% of files. 
\autoref{lst:security} outlines an example of these kinds of instructions.\footnote{\url{https://github.com/amun-ai/hypha/blob/main/CLAUDE.md?plain=1}} 
In this example, the project provides details on the Security Architecture and Permission System. It highlights core principles like Workspace Isolation and Default Protection for preventing unauthorized access, and defines different Service Visibility Levels with their access permissions.









\begin{lstlisting}[style=academicstyle, caption={Example Instruction for Security}, label=lst:security]
## Security Architecture and Permission System
### Core Security Principles
Hypha implements a multi-layered security model with workspace isolation as the foundation. All services have default protection through workspace visibility settings, with additional fine-grained permission controls available for sensitive operations.
### Default Protection: Workspace Isolation
By default, all services are `protected`, meaning they are only accessible to clients within the same workspace. This provides automatic protection against unauthorized cross-workspace access without requiring explicit permission checks in every method.
(continue...)
### Service Visibility Levels
1. `protected` (DEFAULT) - Only accessible by clients in the same workspace
2. `public` - Accessible by all authenticated users across workspaces
3. `unlisted` - Same as public, accessible for all users, but not discoverable
\end{lstlisting}

\par
\smallskip
\textbf{UI/UX. }This kind of instruction contains guidelines or details concerning the user interface (UI) and user experience (UX). This was the second least prevalent category overall, observed in only \cntUIUX\% of the files. 
\autoref{lst:ui} outlines an example of these kinds of instructions.\footnote{\url{https://github.com/bdougie/contributor.info/blob/main/CLAUDE.md}} 
In this example, the project provides User Experience Standards, mandating an ``invisible, Netflix-like user experience'' where data loading and processing occur automatically in the background. The key principles include automatic data handling, subtle notifications, progressive loading, and eliminating the need for manual user intervention.

\begin{lstlisting}[style=academicstyle, caption={Example Instruction for UI/UX}, label=lst:ui]
## User Experience Standards
This project follows an **invisible, Netflix-like user experience** where data loading and processing happens automatically in the background. Key principles:
1. Database-first: Always query cached data before API calls
2. Auto-detection: Automatically detect and fix data quality issues
3. Subtle notifications: Keep users informed without interrupting workflow
4. Progressive enhancement: Core functionality works immediately, enhanced features load in background
5. No manual intervention: Users never need to click "Load Data" or understand technical details

\end{lstlisting}

\par
\smallskip
\textbf{Debugging. }This kind of instruction is one of the new categories added during the analysis because many cases could not be categorized into existing labels. We observed these instructions in \cntDebugging\% of files. 
\autoref{lst:debug} outlines an example of these kinds of instructions.\footnote{\url{https://github.com/probelabs/probe/blob/main/CLAUDE.md}} 
In this example, the project provides Debugging Tips, specifically commands like using \texttt{DEBUG=1} for verbose output, checking \texttt{error.log} for errors, using \texttt{RUST\_BACKTRACE=1} for stack traces, and profiling with \texttt{cargo flamegraph} for performance analysis.

\begin{lstlisting}[style=academicstyle, caption={Example Instruction for Debugging}, label=lst:debug]
### Debugging Tips
- Use `DEBUG=1` for verbose output
- Check `error.log` for detailed errors
- Use `RUST_BACKTRACE=1` for stack traces
- Profile with `cargo flamegraph` for performance
\end{lstlisting}

\begin{table}[!th] 
\small 
\centering 
\caption{Instruction Category Distribution by Agent Type (\%)} 
\label{tab:label_distribution} 
\begin{tabular}{lrrr} 
\toprule 
\textbf{Category} & \textbf{Codex} & \textbf{Claude} & \textbf{Copilot} \\ \midrule 
\lblArchitecture & 10.55\% & 11.20\% & 10.87\% \\ 
\lblBuildRun & 9.23\% & 12.04\% & 8.44\% \\ 
\lblImplDetails & 12.24\% & 10.13\% & 12.16\% \\ 
\lblTest & 13.94\% & 12.16\% & 10.87\% \\ 
\lblDevProcess & 12.62\% & 9.30\% & 10.16\% \\ 
\lblGenSysOverview & 7.72\% & 9.42\% & 10.87\% \\ 
\lblMaintenance & 7.53\% & 7.99\% & 5.87\% \\ 
Configuration \& Env. & 6.40\% & 6.91\% & 5.29\% \\ 
\lblDebugging & 3.01\% & 3.93\% & 5.01\% \\ 
\lblClaudeAIIntegration & 3.39\% & 3.46\% & 4.72\% \\ \bottomrule 
\end{tabular} 
\end{table}

\noindent\textbf{While instruction types are shared, agents show largely consistent distributions 
with minor platform-specific variations.} All three coding agents receive instructions across 
all categories, indicating that developers require similar types of guidance regardless of 
the agent they use. As shown in Table~\ref{tab:label_distribution}, the top categories (\eg \lblTest, \lblImplDetails, 
\lblArchitecture, and \lblBuildRun) remain prominent for all agents, and the bottom categories 
(\eg \lblDebugging, \lblClaudeAIIntegration) are consistently less frequent. The differences in 
percentages across agents are generally small, typically within 1\% to 3\%.

While there are minor variations, our statistical tests confirm that \claude places a 
higher emphasis on \lblBuildRun instructions compared to others (medium effect size, 
$d=-0.34$ vs \codex and $d = -0.36$ vs \copilot), suggesting that \claude is frequently used 
in environments where the agent is expected to actively execute and verify the build. 
Similarly, \codex receives a slightly higher proportion of \lblTest (13.94\%) and 
\lblDevProcess (12.62\%) instructions compared to \copilot or \claude, and \copilot shows a 
higher density of \lblGenSysOverview (10.87\%) and \lblDebugging (5.01\%) instructions compared 
to \codex. However, these differences are not substantial enough to suggest fundamentally 
different usage patterns across agents. Rather, the overall similarity in distributions 
suggests that the nature of developer instructions is primarily driven by the tasks 
themselves, not by the specific agent being used.

\smallskip
\noindent\textbf{Instructions directly govern the structural and stylistic properties of agent-generated code.} Beyond mere categorization, we observe an alignment between the specific constraints defined in the agent context files and the implementation patterns in post-adoption commits. This suggests that agents actively parse and stick to the project-specific rules provided by developers.
For instance, context file instructions often define strict naming conventions and architectural boundaries. Listing~\ref{lst:acf_ex_2} shows an agent context file~\footnote{\url{https://github.com/skhomuti/csm-sentinel/blob/main/AGENTS.md?plain=1}} defining specific Python naming rules and a preference for keeping side effects isolated. Listing~\ref{lst:corrcode_2} demonstrates a corresponding AI-generated commit~\footnote{\url{https://github.com/skhomuti/csm-sentinel/commit/1cd3b568a65b2f06fbddd2aa1e481c6308f07015}} where the agent correctly implemented a batch processing loop using the requested \texttt{UPPER\_SNAKE\_CASE} for environment-derived constants and adhered to the specified logging and indentation styles.


\begin{lstlisting}[style=academicstyle, caption={Example Agent Context File}, label=lst:acf_ex_2]
## Coding Style & Naming
- Python >= 3.11; follow PEP 8; 4-space indentation.
- Names: modules/functions `snake_case`, classes `CapWords`, constants `UPPER_SNAKE_CASE`.
- Prefer type hints and small, focused functions. Keep side effects in orchestrators (`main.py`, `app/bootstrap.py`, `rpc.py`, `jobs.py`).
\end{lstlisting}

\begin{lstlisting}[style=academicstyle, caption={Corresponding Code}, label=lst:corrcode_2]
for batch_start in range(start_block, current_block + 1, batch_size):
    batch_end = min(batch_start + batch_size - 1, current_block)
    for contract in [os.getenv("CSM_ADDRESS"), os.getenv("FEE_DISTRIBUTOR_ADDRESS"), os.getenv("VEBO_ADDRESS"),]:
        logger.debug("Fetching logs for %s blocks %s-%s", contract, batch_start, batch_end,)
\end{lstlisting}

To move beyond the illustrative examples, we conduct two case studies on a simple, machine-checkable rule: 4-space indentation (Listing~\ref{lst:acf_ex_2}). For each case, we measure indentation compliance in AI-authored commits before and after the rule was added to the context file. We focus on this instruction because the intent is unambiguous, compliance can be checked automatically on every added line, and projects that introduce the rule during development provide a natural before-and-after comparison. We identify candidate projects by searching context files for explicit references to 4-space indentation and select one project per agent with a clearly identifiable rule-introduction commit and sufficient AI-authored commits in both periods.
In a repository\footnote{\url{https://github.com/rgerhards/rsyslog}} using \claude, AI-authored commits produced 208 violating lines out of 342 eligible added lines before the rule was introduced, a violation rate of 60.82\%. Afterward, the rate dropped to 11.03\% (939 of 8,516 lines), an absolute reduction of about 50\% despite nearly 25 times more added lines. In a repository\footnote{\url{https://github.com/NB-Core/lotgd}} using \copilot, the violation rate fell from 20.17\% (6,673 of 33,080 lines) before the rule to 7.57\% (5,306 of 70,060 lines) afterward, a reduction of about 13\%. The violation rate decreased substantially in both repositories. Although two cases cannot establish causality, the consistent decline after introducing the rule supports our qualitative observations and suggests that explicit, machine-readable instructions can meaningfully influence code formatting. 



\summarybox{\textbf{Answer to \rqd}}{We identified 16 instruction categories. Instructions for functional aspects (\lblBuildRun, \lblImplDetails, \lblArchitecture, \lblTest) are more prevalent, while meta-level or non-functional categories such as \lblPerformance, \lblSecurity, and \lblUIUX appear far less frequently.
}




\subsection{\rqE}\label{sec:rqe}

\subsubsection{Motivation.}
The manual content analysis in Section~\ref{sec:rqd} provides a detailed taxonomy of instructions in \am, but manual qualitative coding is inherently labor-intensive and not scalable. As the adoption of agentic coding tools accelerates, the volume of \am will grow, making manual inspection a bottleneck for monitoring ecosystem-wide trends.  The motivation for automating this classification is to provide a research instrument that enables the study of agentic coding practices at scale. This automation benefits several key audiences. For researchers, it enables long-term studies to examine how instruction patterns evolve over time and facilitates the comparison of practices across different programming communities or project types without excessive manual effort. It also allows for the easy replication of empirical studies across larger, emerging datasets. For tool builders, understanding the distribution of instructions can inform the design of agent scaffolding tools, linters, or documentation templates that encourage developers to include comprehensive guidance. For practitioners, ecosystem-wide analysis can highlight common blind spots. The goal of this classification is not to feed results back into the LLMs for operational use, but to empower the software engineering community to detect emerging gaps, anti-patterns, and best practices in agent configuration. Recent work suggests that LLMs can effectively perform classification tasks typically reserved for human annotators~\cite{DBLP:conf/msr/AhmedDTP25}. Therefore, we investigate the possibility of automating the classification of agent context files to enable future large-scale monitoring.


\begin{table}[t]
\centering
\caption{Performance of automatic classification using GPT-5 on \am. The highest value within each category is highlighted in \textbf{bold}.}
\label{tab:classification_results}
\begin{tabular}{llrrrr}
\toprule
Category & Label & Precision & Recall & F1-Score & \# Support \\ \midrule
General & System Overview & \textbf{0.88} & \textbf{0.90} & \textbf{0.89} & 196 \\
~ & AI Integration & 0.33 & 0.86 & 0.48 & 80 \\
~ & Documentation and References & 0.43 & 0.87 & 0.57 & 89 \\ \midrule
Implementation & Architecture & \textbf{0.89} & \textbf{0.97} & \textbf{0.93} & 226 \\
~ & Implementation Details & 0.85 & 0.93 & 0.89 & 235 \\ \midrule
Build & Build \& Run & 0.90 & 0.94 & 0.92 & 209 \\
~ & Testing & \textbf{0.91} & \textbf{0.96} & \textbf{0.94} & 252 \\
~ & Configuration \& Environment & 0.64 & 0.91 & 0.75 & 129 \\
~ & DevOps & 0.64 & 0.84 & 0.72 & 61 \\ \midrule
Management & Development Process & \textbf{0.92} & \textbf{0.76} & \textbf{0.83} & 216 \\
~ & Project Management & 0.40 & 0.44 & 0.42 & 18 \\ \midrule
Quality & Maintenance & 0.65 & 0.50 & 0.56 & 148 \\
~ & Debugging & \textbf{0.70} & 0.73 & 0.71 & 84 \\
~ & Performance & 0.59 & 0.90 & 0.71 & 48 \\
~ & Security & 0.60 & \textbf{0.98} & \textbf{0.74} & 49 \\
~ & UI/UX & 0.48 & 0.69 & 0.56 & 29 \\ \midrule
\multicolumn{2}{c}{\textbf{Micro Average}} & 0.73 & 0.86 & 0.79 & 2,069 \\ \bottomrule
\end{tabular}
\end{table}

\subsubsection{Approach.}
We frame the problem as a multi-label classification task over the 16 categories from Section~\ref{sec:rqd} (Table~\ref{tab:category_summary}). We leverage GPT-5 to automatically classify the content of the \numlabelingfiles files from our manually labeled RQ3 dataset, with each request using $temperature = 1$. This dataset comprises a sample of \claude, \codex, and \copilot files, providing a diverse representation of agent context instructions. By using this specific subset, we ensure that the model's performance is measured against the high-quality consensus labels reached by our three human inspectors.

The model is formulated as a multi-label binary classifier. We construct a prompt containing (i) the full \file content and (ii) a concise definition of each category accompanied by representative examples. To make the model's output directly consumable by our evaluation pipeline, we constrain it to produce predictions using a JSON structured-output schema. The prompt and implementation details are available in our replication package.
These examples were selected based on the most frequently recurring patterns identified during the Phase 2 manual labeling in RQ3, rather than chosen at random. For instance, for the Architecture category, we included snippets showing directory mappings as in \autoref{lst:architecture}, as this was the most common structural pattern found in the ground truth.

To evaluate performance, we compare the model's predictions against the human-established ground-truth labels from Section~\ref{sec:rqd}. We compute precision, recall, and F1-score per category, as well as the micro average across all 2,069 label assignments.

\subsubsection{Findings.}
\textbf{Automatic classification of \am is feasible and promising, with an overall micro-average F1-score of 0.79.}
As shown in Table~\ref{tab:classification_results}, the model performs particularly well on concrete, functional instructions in the Implementation, Build, and Management categories. Labels such as \lblGenSysOverview, \lblArchitecture, and \lblTest achieve high F1-scores, indicating that instructions in these areas are distinct and consistently expressed, which allows the model to recognize their patterns reliably. \lblBuildRun, \lblImplDetails, and \lblDevProcess also exhibit strong performance.

In contrast, the model struggles with categories that are more abstract or have semantically overlapping or sparsely represented examples. Within the Quality and General categories, labels such as \lblMaintenance, \lblProjMgmt, \lblClaudeAIIntegration, and \lblDocRefs obtain comparatively lower F1-scores. This suggests that, while the model is effective at detecting concrete commands and technical structures, it has more difficulty distinguishing nuanced or infrequent instructions related to project management or maintainability.

\summarybox{\textbf{Answer to \rqe}}{
Automatic classification of \am is feasible, achieving a micro-average F1-score of 0.79. The model performs well on concrete functional categories such as \lblArchitecture and \lblTest, but shows weaker performance on more abstract or infrequent categories such as \lblMaintenance.
}

\section{Implications}\label{sec:implications}

In this section, we discuss the implications of our findings for researchers, developers, and coding agent builders.

\subsection{Implications for researchers}

\textbf{Define and measure context debt as a new form of technical debt.} 
Our analysis in Section~\ref{sec:rqa} shows that \am are not merely configuration files but extensive documentation artifacts. In particular, \claude files exhibit a median Flesch reading ease (FRE) score of 42.39, which classifies them as difficult to read and comparable to high school-level material under the FRE interpretation bands. This low readability, combined with their considerable length, suggests that as projects evolve, these \files accumulate context debt: a state where instructions meant to clarify context for an AI agent become unmaintainable and opaque to human collaborators. This creates a paradox where a mechanism designed to align agents increases the cognitive load on developers. Future work should formalize context debt by identifying specific \file smells (\eg ambiguous directives, conflicting role definitions) and by developing metrics beyond standard readability scores to assess the maintainability of agent instructions.

\smallskip
\textbf{Investigate the granular impact of structural elements on agent performance.}
Our findings in Section~\ref{sec:rqa} reveal that developers rely on a consistent, shallow hierarchy to organize agent context. It remains an open research question whether removing these H1 and H2 headers, or altering their hierarchy, would significantly degrade an LLM's comprehension or its ability to execute tasks accurately. Recent studies report that agent context files fundamentally alter agent efficiency and repository exploration behavior~\cite{lulla2026agentsmd, gloaguen2026agentsmd}. Future work should move beyond binary presence or absence evaluations and conduct controlled ablation studies on the structural markers themselves. Quantifying how formatting, hierarchical depth, and header removal impact downstream programming tasks will clarify if current developer practices optimize or inadvertently hinder agentic workflows.

\smallskip
\textbf{Model the co-evolution of \am and code to automate maintenance.} 
Contrary to the write-once nature of traditional READMEs~\cite{gaughan2025introductionreadmecontributingfiles}, our findings in Section~\ref{sec:rqc} show that \am are actively maintained, with 67.0\% of \claude files undergoing multiple modifications.  
These updates occur at day-scale intervals (median 23.8 hours for \claude, 68.0 hours for \copilot) and involve incremental additions rather than deletions. 
This active, ongoing maintenance pattern suggests that developers revisit \am as they encounter new project requirements over the course of regular development, which is consistent with a co-evolutionary relationship between \am and the underlying codebase. However, our current data do not directly trace which code changes trigger which \file updates. Future work should analyze paired code-and-\file commits to establish a more direct co-evolutionary link and quantify the degree of coupling. 
Building on such evidence, researchers could develop CI-integrated tools that automatically detect divergences. For example, a linter for \am could verify that the commands listed in the \lblBuildRun section (\cntBuildRun\% prevalence) match the actual scripts in \texttt{package.json} or \texttt{Makefile}, which would reduce the manual effort required for synchronization.

\smallskip
\textbf{Investigate the blind spots of agentic coding and build related benchmarks.} 
Our content analysis (Section~\ref{sec:rqd}) shows that functional categories like \lblBuildRun and \lblImplDetails dominate, while non-functional requirements (NFRs) such as \lblSecurity (\cntSecurity\%) and \lblPerformance (\cntPerformance\%) are notably rare. 

However, we note that the low prevalence of NFR instructions does not necessarily mean that developers neglect these concerns altogether. Developers may address security and performance through other means, such as manual code reviews, external tooling (\eg static analysis, CI/CD security scanners), or by gradually adding NFR-related instructions in later commits. Our current analysis captures the presence of instruction categories at a single snapshot but does not track how the semantic content of individual commits evolves over time. Identifying when developers incorporated NFR instructions would require applying the automatic classification approach from RQ4 to each commit diff, which demands considerable effort and computational resources, and is left as future work. 
Nonetheless, the current gap in agent-facing NFR guidance remains noteworthy. If benchmarks (\eg SWE-bench~\citep{jimenez2024swebench}) do not penalize agents for generating insecure or inefficient code, this blind spot will persist. The research community should design new benchmarks or evaluation frameworks that explicitly test an agent's adherence to NFRs defined in \am. Future work should also examine the post-generation activities of developers to understand whether and how they compensate for the absence of NFR instructions through testing, code review, or complementary tooling.

\subsection{Implications for developers}

\textbf{Adopt a configuration-as-code mindset for \am.}
The active maintenance patterns observed in Section~\ref{sec:rqc} indicate that \am behave more like dynamic configurations than static text. Developers should therefore treat files like \claudemd or \agentsmd with the same rigor applied to \texttt{Dockerfiles} or CI/CD workflows. We recommend integrating \file updates into the standard code review process. When a pull request modifies the build system or refactors a core module, the review checklist should explicitly ask whether the change requires an update to the \am. Treating these files as living code artifacts prevents context drift and keeps the agent's operational knowledge aligned with the current state of the project.

\smallskip
\textbf{Explicitly include non-functional requirements~(NFRs) to prevent quality degradation.}
The scarcity of instructions related to \lblSecurity and \lblPerformance (Section~\ref{sec:rqd}) serves as a warning. 

This challenge may be compounded by the fact that writing effective LLM instructions is inherently difficult, even for motivated users~\cite{DBLP:conf/chi/Zamfirescu-Pereira23}, which could lead developers to focus on familiar functional topics while omitting less tangible quality attributes.

Unlike human senior developers who may implicitly understand secure coding practices, AI agents rely heavily on explicit context. If security guidelines are absent from the \file, agents may produce functional yet vulnerable code (\eg SQL injection risks). Developers should proactively structure their \am to include mandatory sections for NFRs. Directives such as ``All database interactions must use parameterized queries'' or ``Do not commit secrets'' should be codified within the \file to act as continuous guardrails during generation.

\smallskip
\textbf{Implement versioning and governance for \am' evolution.}
Given that \am grow through frequent, small additions (Section~\ref{sec:rqc}), they risk becoming unstructured append-only logs. To address this, developers should apply semantic versioning to their \files and maintain a changelog. Because agents may misinterpret instructions, changes to high-impact sections like \lblArchitecture or \lblDevProcess should require approval from a designated context owner via \texttt{CODEOWNERS},\footnote{\url{https://docs.github.com/en/repositories/managing-your-repositorys-settings-and-features/customizing-your-repository/about-code-owners}} ensuring that the agent's behavioral guidelines remain consistent and authoritative.


\subsection{Implications for coding agent builders}

\textbf{Leverage structural consistency for scaffolded authoring tools.}
Our structural analysis in Section~\ref{sec:rqa} shows that \am follow a consistent, shallow hierarchy based on H1 and H2 headers. Tool builders can use this pattern to lower the barrier to entry for developers. Instead of presenting an empty text editor, IDEs and agent interfaces should provide templated scaffolds pre-populated with the common categories identified in Section~\ref{sec:rqd} (\eg \lblBuildRun, \lblImplDetails, \lblTest). These templates should also include placeholders for frequently neglected NFR categories (\lblSecurity, \lblPerformance), which would encourage more comprehensive documentation.

\textbf{Optimize context retrieval using semantic categorization.}
The categorization of instructions (Table~\ref{tab:category_summary} in Section~\ref{sec:rqd}) indicates that \am contain distinct clusters of information. Retrieval-augmented generation~(RAG) systems for coding agents should move beyond naive text chunking and use this semantic structure. When an agent is tasked with fixing a bug, for example, retrieval should prioritize sections labeled \lblDebugging (\cntDebugging\%) and \lblTest, while deprioritizing unrelated sections like \lblDeployOps. By parsing the hierarchy of an \file, agent builders can improve the signal-to-noise ratio of the context supplied to the LLM, which should lead to more accurate and relevant code generation.

\section{Related Work}\label{sec:related}

In this section, we discuss related work on AI agents in software engineering, software documentation, and context engineering.

\subsection{AI agents in software engineering}

AI in software development is shifting from single-turn code completion to agentic software engineering~\citep{hassan2025agenticse}. In this paradigm, agents understand high-level goals, produce plans, and execute multi-step tasks with minimal human intervention~\citep{Wangetal2024, Bouzeniaetal2024, Schicketal2023, Bubecketal2023}. These agents combine an LLM-based reasoning core with memory, planning modules, and tool use, such as interacting with IDEs, running tests, and invoking external services~\citep{Feldtetal2023, Mialonetal2023}. This reframes LLMs from autocomplete engines into components within larger systems that must be configured, orchestrated, and maintained~\citep{Jinetal2024, Xuetal2022}. Architectures are converging on perception, memory, and action patterns, where agents ingest context, ground decisions in persistent memory, and act through tool APIs~\citep{Wangetal2024}.

Research has expanded from simple code generation to agents that can interactively clarify ambiguous requirements~\citep{Muetal2024, Endresetal2023, wu_huamnevalcomm_2025}, refactor code~\citep{Alomaretal2024, horikawa2025agenticrefactoring}, and assess code quality~\citep{Yedidaetal2022}. Collaboration mechanisms further boost effectiveness. For example, multi-agent pair programming splits strategic planning and tactical coding across coordinated roles, such as Navigator and Driver agents~\citep{Zhangetal_12024}, and scalable human-in-the-loop frameworks that keep developers in control of plans and reviews while letting agents draft and iterate code~\citep{Pasuksmitetal2025, Takerngsaksirietal2024}. Existing studies primarily focus on the agent's internal architecture and collaborative behavior, that is, how the agents reason and coordinate, rather than how they are configured from the outside.

Automatic program repair~(APR) has been an early beneficiary of this paradigm. Base LLMs already rival or surpass traditional APR pipelines on several benchmarks~\citep{Xiaetal2022, Fanetal2022, Gouesetal2019}, while agentized repair further improves reliability by iterating between code changes, test execution, and analysis~\citep{Bouzeniaetal2024}. Related work integrates specification extraction and formal feedback into the loop so that agents can better infer intent and refine patches~\citep{Ruanetal2024, Endresetal2023, Muetal2024}. Beyond repair and generation, agents have been proposed for autonomous testing~\citep{Feldtetal2023} and code quality support, such as detecting smells and providing maintainability guidance~\citep{Yedidaetal2022}.

With greater integration comes new failure modes. Empirical studies report integration defects and security risks across agent, vector store, connector, and execution layers~\citep{Shaoetal2024},
and recent work has examined how developers perceive and intervene on LLM-generated code in real projects~\cite{DBLP:conf/llm4code/RasnayakaWSI24}. 
For example, the Model Context Protocol~(MCP) helps agents integrate with tools but introduces security risks~\citep{hasan2025modelcontextprotocolmcp}. At the same time, evaluating agents remains costly and variable. Comprehensive test-based assessments are expensive, while LLM-as-a-judge evaluations can be inconsistent. Real-world utilization depends on collaboration patterns and process fit~\citep{Pasuksmitetal2025, Takerngsaksirietal2024, Surietal2023}. These observations highlight the need for reliable artifacts and practices that communicate project-specific constraints to agents and keep behavior aligned over time.


Our work targets this missing layer: persistent configuration and context artifacts that steer agent behavior in real-world projects. Prior work on agentic software engineering largely concentrates on agents' internal architectures, tool-use loops, and coordination. When external steering is discussed, it is often framed as prompt engineering, i.e., task-level instructions crafted for a single interaction~\citep{Shinetal2023, Zhangetal2024}. In contrast, we examine project-level documents that configure agents from the outside and are automatically loaded at the start of each session, namely \am such as \claudemd, \agentsmd, and \copilotmd. Rather than being discarded after one task, these context files accumulate roles, constraints, build and test routines, and coding conventions over time, and must be maintained alongside the codebase. While augmented language model designs advocate the use of external tools, memory, and retrieval to ground decisions~\citep{Mialonetal2023}, the community lacks a systematic understanding of how open-source projects encode such persistent guidance for agents at scale. To our knowledge, this is the first empirical analysis of these files, quantifying their structure, maintenance, and instruction taxonomies to complement existing work with evidence about this configuration layer.

\subsection{Software documentation}

Empirical research on software documentation spans a wide range of artifacts, such as READMEs, API documentation, issue discussions, and release notes. Prior studies found that README files are typically concise and function-focused, often expanding over time~\cite{gaughan2025introductionreadmecontributingfiles}. Over 90\% of READMEs mention basic information such as project name, description, and usage instructions~\cite{prana2019categorizing}. Large-scale repository mining has uncovered recurring documentation issues and antipatterns in APIs~\cite{Aghajanietal2018}, fragmented feature documentation in popular GitHub projects~\cite{Puhlfurssetal2022}, and  patterns in how refactoring needs and improvement activities are described and tracked through issues~\cite{Alomaretal2022}. Surveys and interviews further identify what practitioners value and which issues they consider most critical~\cite{Aghajanietal2020, Alsuhaibanietal2021}.

Researchers have reported documentation issues found in software projects~\cite{Aghajanietal2019}, but evaluating documentation quality remains challenging. Automatic metrics adapted from machine translation and summarization, such as BLEU, ROUGE, and METEOR, are widely used to assess generated comments and summaries~\cite{Mastropaoloetal2023, Raietal2022,Guoetal2023}. However, several studies show that improvements in these metrics often fail to align with human-perceived usefulness, with only moderate correlations and even ranking inversions relative to practitioner judgments~\cite{Royetal2021, Kruseetal2024, Stapletonetal2020}. 

Beyond static quality, research also examines how documentation evolves. Studies on code-comment co-evolution show frequent inconsistencies~\cite{Wenetal2019}, and techniques have been developed to detect outdated or contradictory statements in API documentation and narratives~\cite{Tanetal2023,Zhouetal2020}. Dynamic and hybrid approaches leverage runtime values or execution traces to generate or validate documentation, showing benefits for specific method categories~\cite{Suliretal2017,Sulir2018}. Recent datasets explicitly target code-documentation alignment during maintenance, underscoring that synchronization remains a persistent challenge~\cite{Paietal2025}.

Documentation concerns also surface in code review. Large-scale analyses show that reviewers routinely discuss, request, and amend documentation, making review a key gatekeeper for documentation quality~\cite{Raoetal2022}. Automated assistance increasingly participates in this process: review bots influence project throughput and communication patterns~\cite{Wesseletal2020}, and specialized frameworks detect and repair defects in API directives, including parameters and exceptions, with high precision~\cite{Zhouetal2020}.

Prior work has primarily investigated developer-facing documentation, such as READMEs, API references, comments, and review discussions, focusing on its quality, evolution, and automation. In contrast, we study agent-facing, persistent configuration artifacts (\am) that encode project context, roles, and operating rules for coding agents. Because their primary audience is an AI agent, these files often use imperative phrasing and explicit constraints, and they tend to be tightly coupled to day-to-day development workflows, including build, test, and contribution processes, and therefore to code evolution, which we quantify in RQ2. Moreover, our taxonomy of 16 instruction categories (RQ3) surfaces content such as AI integration guidance and explicit role definitions that has little counterpart in traditional documentation studies. To our knowledge, these agent context files have not been examined at scale in the documentation literature cited above.

\subsection{Context engineering}

A growing body of work investigates context engineering, which extends beyond traditional prompt engineering to optimize how developers provide information to AI coding agents. While prompt engineering typically optimizes transient, task-level interactions, context engineering concerns persistent artifacts such as configuration files, documentation, and contextual instructions that continuously shape agent behavior across sessions and projects. Researchers are empirically examining how different instruction strategies affect agent performance on diverse software engineering tasks, including code generation \cite{Shinetal2023}, automated program repair \cite{Xiaetal2022, Yeetal2023}, and repository-level code completion \cite{Shrivastavaetal2022}.

This research can be broadly divided into two complementary areas: (a) the study of immediate, task-specific prompts, and (b) the study of persistent, project-level context. Much of the existing literature focuses on optimizing the immediate instructions that developers provide for specific, in-the-moment tasks. These include reasoning-oriented techniques like chain-of-thought (CoT) for multi-step tasks \cite{Pengetal2023}, interactive methods involving reflection and dialogue \cite{Zhangetal2024, Mondaletal2024}, and analyses of recurring patterns in developers' conversational prompts \cite{kumar2025sharptoolsdeveloperswield}. Other studies examine knowledge-enriched prompts that embed domain-specific data, such as UML model constraints \cite{Abukhalafetal2023} or exception handling patterns \cite{Renetal2023}, to produce more robust and reliable outputs.

Complementing this, other work highlights that a primary failure mode for AI assistants is a fundamental ``lack of contextual awareness''~\cite{akhoroz2025conversationalaicodingassistant, tufano2024autodevautomatedaidrivendevelopment}, a problem that cannot be solved by a single prompt alone. This challenge has driven the development of mechanisms that provide agents with persistent, repository-level context~\cite{hai2025impactscontextsrepositorylevelcode, Shrivastavaetal2022}. These context-rich approaches integrate repository-level information such as imports, module hierarchies, or architectural metadata to improve code completion and adaptation \cite{Gaoetal2024, Shrivastavaetal2022}. Studies consistently show that such fine-grained, knowledge-driven context yields substantial improvements in code quality, correctness, and reliability over simple zero-shot prompts~\cite{Shinetal2023, Abukhalafetal2023, Nashidetal2023}. As providing this context becomes a core development activity, managing these instructions as long-lived artifacts has emerged as a new software engineering challenge \cite{li2025promptgithub}. 

While prior studies have explored how different prompt- and retrieval-based context strategies affect downstream task quality~\cite{Shrivastavaetal2022, Gaoetal2024, kumar2025sharptoolsdeveloperswield}, they typically treat context as an experimental input and evaluate its \emph{effect} on model performance. Our work is complementary. Rather than measuring the effect of providing context, we study the long-lived artifacts that practitioners create to provide and manage that context in real repositories. This paper presents the first empirical study of \am (\eg \claudemd, \agentsmd) as key artifacts of context engineering, analyzing their structure, content, and maintenance patterns to understand how developers curate the persistent instructions that guide agentic coding in practice.

\section{Threats to Validity}\label{sec:threats_to_validity}

In this section, we discuss the threats to validity of our study about \am.

\subsection{Internal Validity}
Threats to internal validity concern factors internal to our study that could have influenced our results. Section~\ref{sec:rqd} involves manual classification of content within the \am, which introduces potential human error and subjective bias. To address this risk, two inspectors examined the \am independently and carefully. This independent labeling achieved \agreementRate agreement, and a third inspector resolved any conflicting labels.

\subsection{Construct Validity}
Threats to construct validity concern the extent to which the measures used truly reflect the concepts being studied. Our manual content classification approach categorized instructions into 16 distinct labels based on the presence of a topic. For instance, the category Implementation Details encompasses specific guidance for implementing code or system components, including coding style guidelines. The threat arises because this classification was purely binary; an agent context file was flagged with the Implementation Details label if it contained any mention of code style, regardless of whether that content was minimal (one line of instruction) or substantial (many lines detailing complex conventions). Consequently, the frequency reported for a category represents only the prevalence of the topic, not the depth, complexity, or qualitative richness that developers invested in that specific instruction set.

To confirm the reliability of our programmatic approach for assessing readability in technical artifacts, we cross-checked our findings using an alternative framework, py-readability-metrics\footnote{\url{https://pypi.org/project/py-readability-metrics/}}. Across the studied agent context files, the Mann-Whitney U test confirms no significant difference between the tools across the studied files (median difference of 0.279), demonstrating the robustness of Textstat. Manual inspection of the minor variations showed that they stem from expected differences in parsing non-prose elements. Textstat treats newlines and bullet points as sentence boundaries, appropriately capturing the structure of list-heavy technical files, whereas py-readability-metrics strictly requires terminal punctuation. Additionally, slight fluctuations arise from differences in syllable estimation techniques for technical strings such as file paths. Overall, these parsing differences are marginal and do not compromise the reliability of Textstat.

Another construct validity concern applies to the use of FRE as a readability proxy for agent context files. FRE was designed and validated for general English prose, and its two underlying assumptions, that longer sentences and higher syllable counts per word indicate greater comprehension difficulty, do not map cleanly onto technical, procedural, and code-adjacent text. As our manual inspection revealed, FRE scores are strongly influenced by content form rather than true cognitive difficulty for the intended audience. Files composed of short imperative directives and inline command tokens receive high (easy) FRE scores because their surface-level lexical and syntactic properties mimic simple prose, yet they may be no easier to act upon for a developer unfamiliar with the codebase. Conversely, files providing rich architectural context with stacked technical noun phrases, framework names, and acronym expansions receive low (hard) FRE scores, even though experienced developers in the relevant domain may find such descriptions clear and actionable. Consequently, low FRE scores in our dataset may partly reflect technical vocabulary and document form rather than genuine comprehension difficulty. Researchers and developers should therefore treat FRE scores in this context as a measure of surface-level lexical and syntactic complexity rather than as a direct measure of usability or clarity for the developers who use these files.

A further concern relates to our dependence on Markdown header syntax as the basis for measuring the structural organization of agent context files. Markdown is flexible, and developers may use its structural elements inconsistently across projects. For example, some developers may use bold text or horizontal rules to define each section rather than explicit header tags, while others may use deeper header levels (e.g., H3, H4) to indicate what others would express as bullet points or paragraph emphasis. Our header-counting approach treats only lines beginning with one or more hash symbols followed by a space as headers, which means that visually structured but syntactically unmarked sections are not captured in our analysis. Consequently, the reported distributions of H1 through H5 headers reflect explicit Markdown usage patterns rather than the full range of structural conventions developers may employ. We chose this strict syntactic definition because it is standardized across all agent context files in our dataset, whereas inferring structure from visual or semantic cues would introduce subjective interpretation. Nonetheless, readers should interpret our structural findings as characterizing how developers use formal Markdown headers, rather than as a complete account of all organizational strategies present in agent context files. Future work could complement our analysis by examining alternative structural signals such as bolded section titles, horizontal rule separators, or list-based hierarchies.

\subsection{External Validity}
Threats to external validity concern the generalizability of our findings. This study examined \numfiles \am from \numrepos repositories that use one of three major agentic coding systems (\ie \claude, \codex, and \copilot). Although we expanded the number of studied files compared to our previous study~\cite{chatlatanagulchai2025agenticmanifests}, the dataset remains limited, which constrains the generalizability of our findings. Future work should extend this analysis to include a larger number of files from various agentic coding systems.

\section{Conclusion}\label{sec:conclusion}
This study presented the first empirical analysis of the structure, maintenance, and content of \am, such as \claudemd, \agentsmd, and \copilotmd, which play an important role in defining and operationalizing AI agent behavior in agentic coding. The primary motivation for this research was the significant lack of accessible documentation for creating these files, which has historically forced developers into inefficient trial-and-error approaches when configuring their agents. Our investigation systematically analyzed \numfiles context files collected from \numrepos open-source repositories to provide an empirical foundation for best practices.

Our findings regarding the characteristics of \am reveal that these files are generally long and difficult to read. Additionally, such files for Claude Code and GitHub Copilot are substantially longer than those for OpenAI Codex, suggesting developers provide a much larger volume of natural language instructions to these agents. Critically, the readability of these files is poor; the documents are complex, with many categorized as the ``difficult'' category comparable to high school-level material. Structurally, \am are consistently organized with a shallow hierarchy, anchored by a single H1 heading and using H2 and H3 subsections to define major topics, a pattern that likely aids developers in quickly parsing and maintaining the documents. Furthermore, unlike conventional software documentation often characterized as ``write-once,'' \am are actively maintained, behaving as evolving configuration artifacts. The majority of \claude files (67.0\%) are revisited and refined across multiple commits, with maintenance occurring in short, rapid bursts. This evolution is driven by small, incremental content additions, while content deletions are negligible.

The analysis of instructional content demonstrated that developers primarily focus on action-oriented, functional guidance necessary for execution and maintenance. The most prevalent instructions address practical operations, including Testing (\cntTest\%), Implementation Details (\cntImplDetails\%), and Architecture (\cntArchitecture\%). However, the study identified critical omissions: instructions addressing non-functional requirements (NFRs), such as Security (\cntSecurity\%), Performance (\cntPerformance\%), and UI/UX (\cntUIUX\%), were notably infrequent. This suggests that agents are extensively guided on ``how'' to build the code functionally, but not necessarily on ``how to build it well'' concerning quality attributes.

\begin{acks}
We gratefully acknowledge the financial support of JSPS KAKENHI grants (JP24K02921, JP25K03100, JP26H02500), as well as ASPIRE grant (JPMJAP2415), JST CREST grant (JPMJCR23M1, JPMJCR26X7), and JST FOREST (JPMJFR252W). We also acknowledge the support of the Natural Sciences and Engineering Research Council of Canada (NSERC).
\end{acks}

\balance
\bibliographystyle{ACM-Reference-Format}
\bibliography{references}

@article{tufano2024autodevautomatedaidrivendevelopment,
  author = {Michele Tufano and Anisha Agarwal and Jinu Jang and Roshanak Z. Moghaddam and Neel Sundaresan},
  title  = {AutoDev: Automated AI-Driven Development},
  journal      = {CoRR},
  volume       = {abs/2403.08299},
  year         = {2024},
  doi          = {10.48550/ARXIV.2403.08299}
}

@article{kumar2025sharptoolsdeveloperswield,
      author       = {Aayush Kumar and
                  Yasharth Bajpai and
                  Sumit Gulwani and
                  Gustavo Soares and
                  Emerson R. Murphy{-}Hill},
      title        = {Sharp Tools: How Developers Wield Agentic {AI} in Real Software Engineering Tasks},
      journal      = {CoRR},
      volume       = {abs/2506.12347},
      year         = {2025},
      doi          = {10.48550/ARXIV.2506.12347},
}

@inproceedings{gaughan2025introductionreadmecontributingfiles,
      author       = {Matthew Gaughan and
                  Kaylea Champion and
                  Sohyeon Hwang and
                  Aaron Shaw},
      title        = {The Introduction of {README} and {CONTRIBUTING} Files in Open Source Software Development},
      booktitle    = {Proceedings of the 2025 ACM/IEEE International Conference on Connected Health: Applications, Systems and Engineering Technologies (CHASE'25)},
      pages        = {191--202},
      year         = {2025},
      doi          = {10.1109/CHASE66643.2025.00030},
}

@article{prana2019categorizing,
  author    = {Prana, Gilang A. A. and Treude, Christoph and Thung, Ferdian and Lo, David and Jiang, Lingxiao},
  title     = {Categorizing the Content of GitHub README Files},
  journal   = {Empirical Software Engineering},
  volume    = {24},
  number    = {3},
  pages     = {1296--1327},
  year      = {2019},
  doi       = {10.1007/S10664-018-9660-3},
}

@ARTICLE{7000568,
  author={Treude, Christoph and Robillard, Martin P. and Dagenais, Barthélémy},
  journal={{IEEE} Transactions on Software Engineering}, 
  title={Extracting Development Tasks to Navigate Software Documentation}, 
  year={2015},
  volume={41},
  number={6},
  pages={565-581},
    doi          = {10.1109/TSE.2014.2387172}
}

@article{gao2025adaptinginstallationinstructionsrapidly,
      author       = {Haoyu Gao and
                  Christoph Treude and
                  Mansooreh Zahedi},
      title        = {Adapting Installation Instructions in Rapidly Evolving Software Ecosystems},
      journal      = {{IEEE} Transactions on Software Engineering},
      volume       = {51},
      number       = {4},
      pages        = {1334--1357},
      year         = {2025},
      doi          = {10.1109/TSE.2025.3552614}
}

@inproceedings{ferreira2021inside_commits,
    author       = {M{\'{\i}}vian M. Ferreira and
                  Diego Gon{\c{c}}alves and
                  Kecia Aline M. Ferreira and
                  Mariza A. S. Bigonha},
      title        = {Inside Commits: An Empirical Study on Commits in Open-Source Software},
      booktitle    = {Proceedings of the 35th Brazilian Symposium on Software Engineering},
      pages        = {11--15},
      year         = {2021},
      doi          = {10.1145/3474624.3474629}
}

@inproceedings{hai2025impactscontextsrepositorylevelcode,
      author       = {Nam Le Hai and
                  Dung Manh Nguyen and
                  Nghi D. Q. Bui},
      title        = {On the Impacts of Contexts on Repository-Level Code Generation},
      booktitle    = {Proceedings of the Findings of the Association for Computational Linguistics (NAACL'25)},
      pages        = {1496--1524},
      year         = {2025},
        doi          = {10.18653/V1/2025.FINDINGS-NAACL.82}
}

@inproceedings{min2022rethinking,
  title = {Rethinking the Role of Demonstrations: What Makes In-Context Learning Work?},
  author = {Min, Sewon and Lyu, Xinxi and Holtzman, Ari and Artetxe, Mikel and Lewis, Mike and Hajishirzi, Hannaneh and Zettlemoyer, Luke},
  booktitle    = {Proceedings of the 2022 Conference on Empirical Methods in Natural
                  Language Processing ({EMNLP}'22)},
  pages        = {11048--11064},
  year         = {2022},
  doi          = {10.18653/V1/2022.EMNLP-MAIN.759},
}

@inproceedings{DBLP:conf/msr/AhmedDTP25,
  author       = {Toufique Ahmed and
                  Premkumar T. Devanbu and
                  Christoph Treude and
                  Michael Pradel},
  title        = {Can LLMs Replace Manual Annotation of Software Engineering Artifacts?},
  booktitle    = {Proceedings of the 22nd {IEEE/ACM} International Conference on Mining Software Repositories (MSR'25)},
  pages        = {526--538},
  year         = {2025},
  doi          = {10.1109/MSR66628.2025.00086}
}

@inproceedings{chatlatanagulchai2025agenticmanifests,
      title={On the Use of Agentic Coding Manifests: An Empirical Study of Claude Code}, 
      author={Worawalan Chatlatanagulchai and Kundjanasith Thonglek and Brittany Reid and Yutaro Kashiwa and Pattara Leelaprute and Arnon Rungsawang and Bundit Manaskasemsak and Hajimu Iida},
      year={2025},
    booktitle    = {Proceedings of the 26th International Conference on Product-Focused Software Process Improvement (PROFES'25)},
  doi={10.1007/978-3-032-12089-2_40}
}

@article{Mann1947OnAT,
	title        = {On a Test of Whether one of Two Random Variables is Stochastically Larger than the Other},
	author       = {H. B. Mann and D. R. Whitney},
	year         = 1947,
	journal      = {Annals of Mathematical Statistics},
	volume       = 18,
	pages        = {50--60},
    doi          = {10.1214/aoms/1177730491}
}

@incollection{Cliff,
	title        = {Ordinal {Analysis} of {Behavioral} {Data}},
	author       = {Long, Jeffrey D. and Feng, Du and Cliff, Norman},
	year         = 2003,
	month        = apr,
	booktitle    = {Handbook of {Psychology}},
	publisher    = {John Wiley \& Sons, Inc.},
	address      = {Hoboken, NJ, USA},
	pages        = {635--661},
	editor       = {Weiner, Irving B.},
	chapter      = 25,
    doi       = {10.1002/0471264385.wei0225}
}

@inproceedings{Cliff_threshold,
	title        = {Exploring methods for evaluating group differences on the {NSSE} and other surveys: Are the t-test and {C}ohen’sd indices the most appropriate choices},
	author       = {Romano, Jeanine and Kromrey, Jeffrey D and Coraggio, Jesse and Skowronek, Jeff and Devine, Linda},
	year         = 2006,
	booktitle    = {annual meeting of the Southern Association for Institutional Research},
	pages        = {1--51},
	organization = {Citeseer}
}

@article{hassan2025agenticse,
      title={Agentic Software Engineering: Foundational Pillars and a Research Roadmap}, 
      author={Ahmed E. Hassan and Hao Li and Dayi Lin and Bram Adams and Tse-Hsun Chen and Yutaro Kashiwa and Dong Qiu},
      year={2025},
      eprint={2509.06216},
      archivePrefix={arXiv},
      primaryClass={cs.SE},
      doi={10.48550/ARXIV.2509.06216},
}

@misc{li2025promptgithub,
      title={Understanding Prompt Management in GitHub Repositories: A Call for Best Practices}, 
      author={Hao Li and Hicham Masri and Filipe R. Cogo and Abdul Ali Bangash and Bram Adams and Ahmed E. Hassan},
      year={2025},
      eprint={2509.12421},
      archivePrefix={arXiv},
      primaryClass={cs.SE},
      doi={10.48550/ARXIV.2509.12421},
}

@article{piantadosi2020readability,
author = {Piantadosi, Valentina and Fierro, Fabiana and Scalabrino, Simone and Serebrenik, Alexander and Oliveto, Rocco},
title = {How does code readability change during software evolution?},
year = {2020},
volume = {25},
number = {6},
journal = {Empirical Softw. Engg.},
month = nov,
pages = {5374--5412},
doi          = {10.1007/S10664-020-09886-9},
}

@INPROCEEDINGS{ehsani2025detectingpromptknowledgegaps,
  author={Ehsani, Ramtin and Pathak, Sakshi and Chatterjee, Preetha},
  booktitle={Proceedings of the 2025 IEEE/ACM 22nd International Conference on Mining Software Repositories (MSR'25)}, 
  title={Towards Detecting Prompt Knowledge Gaps for Improved LLM-guided Issue Resolution}, 
  year={2025},
  volume={},
  number={},
  pages={699--711},
  doi          = {10.1109/MSR66628.2025.00107}
}

@misc{liu2025effectspromptlengthdomainspecific,
      title={Effects of Prompt Length on Domain-specific Tasks for Large Language Models}, 
      author={Qibang Liu and Wenzhe Wang and Jeffrey Willard},
      year={2025},
      eprint={2502.14255},
      archivePrefix={arXiv},
      primaryClass={cs.CL},
      url={https://arxiv.org/abs/2502.14255},
      doi={10.48550/ARXIV.2502.14255},
}

@article{canizares2023readability,
author = {Ca\~{n}izares, Pablo C. and L\'{o}pez-Morales, Jose Mar\'{\i}a and P\'{e}rez-Soler, Sara and Guerra, Esther and de Lara, Juan},
title = {Measuring and Clustering Heterogeneous Chatbot Designs},
year = {2024},
volume = {33},
number = {4},
journal = {ACM Transactions on Software Engineering and Methodology},
articleno = {90},
  doi          = {10.1145/3637228}
}

@article{horikawa2025agenticrefactoring,
      title={Agentic Refactoring: An Empirical Study of AI Coding Agents}, 
      author={Kosei Horikawa and Hao Li and Yutaro Kashiwa and Bram Adams and Hajimu Iida and Ahmed E. Hassan},
      year={2025},
      eprint={2511.04824},
      archivePrefix={arXiv},
      primaryClass={cs.SE},
      url={https://arxiv.org/abs/2511.04824},
      doi={10.48550/ARXIV.2511.04824},
}

@inproceedings{Alomaretal2024,
 author = {Eman Abdullah AlOmar and
                  Anushkrishna Venkatakrishnan and
                  Mohamed Wiem Mkaouer and
                  Christian D. Newman and
                  Ali Ouni},
 booktitle = {Proceedings of the 21st {IEEE/ACM} International Conference on Mining Software Repositories (MSR'24)},
 title = {How to Refactor this Code? An Exploratory Study on Developer-ChatGPT Refactoring Conversations},
 year = {2024},
  doi          = {10.1145/3643991.3645081}
}

@inproceedings{Bouzeniaetal2024,
 author = {Islem Bouzenia and Prem Devanbu and Michael Pradel},
 booktitle = {Proceedings of the 47th {IEEE/ACM} International Conference on Software Engineering (ICSE'24)},
 title = {RepairAgent: An Autonomous, LLM-Based Agent for Program Repair},
 year = {2024},
  doi          = {10.1109/ICSE55347.2025.00157}
}

@inproceedings{Endresetal2023,
 author = {Madeline Endres and Sarah Fakhoury and Saikat Chakraborty and Shuvendu K. Lahiri},
 booktitle = {Proceedings of the ACM International Conference on the Foundations of Software Engineering (FSE'24)},
 title = {Can Large Language Models Transform Natural Language Intent into Formal Method Postconditions?},
 year = {2023},
  doi          = {10.1145/3660791}
}

@inproceedings{Fanetal2022,
 author = {Zhiyu Fan and Xiang Gao and M. Mirchev and Abhik Roychoudhury and Shin Hwei Tan},
 booktitle = {Proceedings of the 45th {IEEE/ACM} International Conference on Software Engineering (ICSE'22)},
 title = {Automated Repair of Programs from Large Language Models},
 year = {2022},
  doi          = {10.1109/ICSE48619.2023.00128}
}

@inproceedings{Feldtetal2023,
 author = {R. Feldt and Sungmin Kang and Juyeon Yoon and Shin Yoo},
 booktitle = {Proceedings of the 38th {IEEE/ACM} International Conference on Automated Software Engineering},
 title = {Towards Autonomous Testing Agents via Conversational Large Language Models},
 year = {2023},
  doi          = {10.1109/ASE56229.2023.00148}
}

@article{Gouesetal2019,
 author = {Claire Le Goues and Michael Pradel and Abhik Roychoudhury},
 journal = {Communications of the ACM},
 title = {Automated program repair},
 year = {2019},
 doi={10.1145/3318162}
}

@inproceedings{Jinetal2024,
 author = {Kailun Jin and Chung-Yu Wang and Hung Viet Pham and Hadi Hemmati},
 booktitle = {Proceedings of the 2024 IEEE International Conference on Software Maintenance and Evolution (ICSME'24)},
 title = {Can ChatGPT Support Developers? An Empirical Evaluation of Large Language Models for Code Generation},
 year = {2024},
 doi          = {10.1145/3643991.3645074},
}

@article{Mialonetal2023,
  author       = {Gr{\'{e}}goire Mialon and
                  Roberto Dess{\`{\i}} and
                  Maria Lomeli and
                  Christoforos Nalmpantis and
                  Ramakanth Pasunuru and
                  Roberta Raileanu and
                  Baptiste Rozi{\`{e}}re and
                  Timo Schick and
                  Jane Dwivedi{-}Yu and
                  Asli Celikyilmaz and
                  Edouard Grave and
                  Yann LeCun and
                  Thomas Scialom},
 journal = {Trans. Mach. Learn. Res.},
 title = {Augmented Language Models: a Survey},
 year = {2023},
}

@article{Muetal2024,
 author = {Fangwen Mu and Lin Shi and Song Wang and Zhuohao Yu and Binquan Zhang and ChenXue Wang and Shichao Liu and Qing Wang},
 journal = {Proceedings of the 2024 ACM International Conference on the Foundations of Software Engineering (FSE'24)},
 title = {ClarifyGPT: A Framework for Enhancing LLM-Based Code Generation via Requirements Clarification},
 year = {2024},
}

@inproceedings{Pasuksmitetal2025,
 author = {Jirat Pasuksmit and Wannita Takerngsaksiri and Patanamon Thongtanunam and C. Tantithamthavorn and Ruixiong Zhang and Shiyan Wang and Fan Jiang and Jing Li and Evan Cook and Kun Chen and Ming Wu},
 booktitle = {Proceedings of the 22nd {IEEE/ACM} International Conference on Mining Software Repositories (MSR'25)},
 title = {Human-In-The-Loop Software Development Agents: Challenges and Future Directions},
 year = {2025},
  doi          = {10.1109/MSR66628.2025.00112},
}

@article{Ruanetal2024,
 author = {Haifeng Ruan and Yuntong Zhang and Abhik Roychoudhury},
 journal = {Proceedings of the 47th {IEEE/ACM} International Conference on Software Engineering (ICSE'24)},
 title = {SpecRover: Code Intent Extraction via LLMs},
 year = {2024},
   doi          = {10.1109/ICSE55347.2025.00080},

}

@inproceedings{Schicketal2023,
 author = {Timo Schick and Jane Dwivedi-Yu and Roberto Dessì and R. Raileanu and M. Lomeli and Luke Zettlemoyer and Nicola Cancedda and Thomas Scialom},
 booktitle = {Proceedings of the Advances in Neural Information Processing Systems 36: Annual Conference on Neural Information Processing Systems (NeurIPS'23)},
 title = {Toolformer: Language Models Can Teach Themselves to Use Tools},
 year = {2023},
}

@article{Shaoetal2024,
 author = {Yuchen Shao and Yuheng Huang and Jiawei Shen and Lei Ma and Ting Su and Chengcheng Wan},
 journal = {Proceedings of the IEEE/ACM 47th International Conference on Software Engineering (ICSE'24)},
 title = {Are LLMs Correctly Integrated into Software Systems?},
 year = {2024},
 doi          = {10.1109/ICSE55347.2025.00204},
}

@inproceedings{Surietal2023,
 author = {Samdyuti Suri and Sankar Narayan Das and Kapil Singi and Kuntal Dey and V. Sharma and Vikrant S. Kaulgud},
 booktitle = {Proceedings of the 38th IEEE/ACM International Conference on Automated Software Engineering (ASE'23)},
 title = {Software Engineering Using Autonomous Agents: Are We There Yet?},
 year = {2023},
 doi          = {10.1109/ASE56229.2023.00174},
}

@article{Takerngsaksirietal2024,
 author = {Wannita Takerngsaksiri and Jirat Pasuksmit and Patanamon Thongtanunam and C. Tantithamthavorn and Ruixiong Zhang and Fan Jiang and Jing Li and Evan Cook and Kun Chen and Ming Wu},
 journal = {2025 IEEE/ACM 47th International Conference on Software Engineering: Software Engineering in Practice (ICSE-SEIP)},
 title = {Human-In-The-Loop Software Development Agents},
 year = {2024},
 doi          = {10.1109/ICSE-SEIP66354.2025.00036},
}

@inproceedings{Xuetal2022,
 author = {Frank F. Xu and Uri Alon and Graham Neubig and Vincent J. Hellendoorn},
 booktitle = {Proceedings of the 6th {ACM} {SIGPLAN} International Symposium on Machine
                  Programming (MAPS'22)},
 title = {A systematic evaluation of large language models of code},
 year = {2022},
  doi          = {10.1145/3520312.3534862},
}

@inproceedings{Yedidaetal2022,
  author       = {Rahul Yedida and
                  Tim Menzies},
  title        = {How to Improve Deep Learning for Software Analytics (a case study
                  with code smell detection)},
  booktitle    = {Proceedings of the 19th {IEEE/ACM} International Conference on Mining Software Repositories ({MSR}'22)},
  pages        = {156--166},
  year         = {2022},
  url          = {https://doi.org/10.1145/3524842.3528458},
  doi          = {10.1145/3524842.3528458}
}

@inproceedings{Zhangetal_12024,
 author = {Huan Zhang and Wei Cheng and Yuhan Wu and Wei Hu},
 booktitle = {Proceedings of the 39th {IEEE/ACM} International Conference on Automated
                  Software Engineering ({ASE}'24)},
 title = {A Pair Programming Framework for Code Generation via Multi-Plan Exploration and Feedback-Driven Refinement},
 year = {2024},
  doi          = {10.1145/3691620.3695506},
}

@article{wu_huamnevalcomm_2025,
author = {Wu, Jie JW and Fard, Fatemeh H.},
title = {HumanEvalComm: Benchmarking the Communication Competence of Code Generation for LLMs and LLM Agents},
year = {2025},
volume = {34},
number = {7},
journal = {ACM Transactions on Software Engineering and Methodology},
articleno = {189},
doi       = {10.1145/3715109},
}

@article{hasan2025modelcontextprotocolmcp,
      title={Model Context Protocol (MCP) at First Glance: Studying the Security and Maintainability of MCP Servers}, 
      author={Mohammed Mehedi Hasan and Hao Li and Emad Fallahzadeh and Gopi Krishnan Rajbahadur and Bram Adams and Ahmed E. Hassan},
      year={2025},
      eprint={2506.13538},
      archivePrefix={arXiv},
      primaryClass={cs.SE},
}

@inproceedings{Aghajanietal2018,
	title        = {A Large-Scale Empirical Study on Linguistic Antipatterns Affecting APIs},
	author       = {Emad Aghajani and Csaba Nagy and G. Bavota and Michele Lanza},
	year         = 2018,
	booktitle      = {Proceedings of the 2018 IEEE International Conference on Software Maintenance and Evolution (ICSME'18)},
    doi          = {10.1109/ICSME.2018.00012}
}

@inproceedings{Aghajanietal2019,
	title        = {Software Documentation Issues Unveiled},
	author       = {Emad Aghajani and Csaba Nagy and Olga Lucero Vega-Márquez and Mario Linares-Vásquez and Laura Moreno and G. Bavota and Michele Lanza},
	year         = 2019,
	booktitle      = {Proceedings of the 42nd International Conference on Software Engineering (ICSE'19)},
  doi          = {10.1109/ICSE.2019.00122}
}

@inproceedings{Aghajanietal2020,
	title        = {Software Documentation: The Practitioners' Perspective},
	author       = {Emad Aghajani and Csaba Nagy and Mario Linares-Vásquez and Laura Moreno and G. Bavota and Michele Lanza and D. Shepherd},
	year         = 2020,
		booktitle      = {Proceedings of the 42nd International Conference on Software Engineering (ICSE'20)},
      doi          = {10.1145/3377811.3380405}
}

@inproceedings{Alomaretal2022,
	title        = {An Exploratory Study on Refactoring Documentation in Issues Handling},
	author       = {Eman Abdullah AlOmar and Anthony Peruma and Mohamed Wiem Mkaouer and Christian D. Newman and Ali Ouni},
	year         = 2022,
	booktitle      = {Proceedings of the 19th {IEEE/ACM} International Conference on Mining Software Repositories (MSR'22)},
  doi          = {10.1145/3524842.3528525}
}

@inproceedings{Alsuhaibanietal2021,
	title        = {On the Naming of Methods: A Survey of Professional Developers},
	author       = {Reem S. Alsuhaibani and Christian D. Newman and M. J. Decker and Michael L. Collard and Jonathan I. Maletic},
	year         = 2021,
	booktitle      = {Proceedings of the 43rd {IEEE/ACM} International Conference on Software Engineering (ICSE'21)},
  doi          = {10.1109/ICSE43902.2021.00061}
}

@article{Guoetal2023,
	title        = {Snippet Comment Generation Based on Code Context Expansion},
	author       = {Hanyang Guo and Xiangping Chen and Yuan Huang and Yanlin Wang and Xi Ding and Zibin Zheng and Xiaocong Zhou and Hong-ning Dai},
	year         = 2023,
	journal      = {ACM Transactions on Software Engineering and Methodology},
    doi          = {10.1145/3611664},
}

@inproceedings{Kruseetal2024,
	title        = {Can Developers Prompt? A Controlled Experiment for Code Documentation Generation},
	author       = {Hans-Alexander Kruse and Tim Puhlfürß and Walid Maalej},
	year         = 2024,
	booktitle      = {Proceedings of the 2024 IEEE International Conference on Software Maintenance and Evolution (ICSME'24)},
    doi          = {10.1109/ICSME58944.2024.00058},

}

@inproceedings{Mastropaoloetal2023,
	title        = {Evaluating Code Summarization Techniques: A New Metric and an Empirical Characterization},
	author       = {Antonio Mastropaolo and
                  Matteo Ciniselli and
                  Massimiliano Di Penta and
                  Gabriele Bavota},
  year         = {2024},
	booktitle      = {Proceedings of the 46th {IEEE/ACM} International Conference on Software
                  Engineering (ICSE'24)},
  doi          = {10.1145/3597503.3639174},
}

@inproceedings{Paietal2025,
	title        = {CoDocBench: A Dataset for Code-Documentation Alignment in Software Maintenance},
	author       = {K. Pai and Prem Devanbu and Toufique Ahmed},
	year         = 2025,
	booktitle      = {Proceedings of the 2025 IEEE/ACM 22nd International Conference on Mining Software Repositories (MSR'25)},
    doi          = {10.1109/MSR66628.2025.00077},
}

@inproceedings{Puhlfurssetal2022,
	title        = {An Exploratory Study of Documentation Strategies for Product Features in Popular GitHub Projects},
	author       = {Tim Puhlfurss and Lloyd Montgomery and W. Maalej},
	year         = 2022,
	booktitle      = {Proceedings of the 2022 IEEE International Conference on Software Maintenance and Evolution (ICSME'22)},
    doi          = {10.1109/ICSME55016.2022.00043},
}

@article{Raietal2022,
	title        = {A Review on Source Code Documentation},
	author       = {Sawan Rai and R. Belwal and Atul Gupta},
	year         = 2022,
	booktitle      = {ACM Transactions on Intelligent Systems and Technology},
    doi       = {10.1145/3519312},
}

@inproceedings{Raoetal2022,
	title        = {Comments on Comments: Where Code Review and Documentation Meet},
	author       = {N. Rao and Jason Tsay and Martin Hirzel and Vincent J. Hellendoorn},
	year         = 2022,
	booktitle      = {Proceedings of the 19th International Conference on Mining Software Repositories (MSR'22)},
    doi          = {10.1145/3524842.3528475},
}

@inproceedings{Royetal2021,
	title        = {Reassessing automatic evaluation metrics for code summarization tasks},
	author       = {Devjeet Roy and Sarah Fakhoury and Venera Arnaoudova},
	year         = 2021,
	booktitle      = {Proceedings of the 29th ACM Joint Meeting on European Software Engineering Conference and Symposium on the Foundations of Software Engineering (FSE'21)},
    doi          = {10.1145/3468264.3468588},
}

@inproceedings{Stapletonetal2020,
	title        = {A Human Study of Comprehension and Code Summarization},
	author       = {Sean Stapleton and Yashmeet Gambhir and Alexander LeClair and Zachary Eberhart and Westley Weimer and Kevin Leach and Yu Huang},
	year         = 2020,
	booktitle      = {Proceedings of the 28th International Conference on Program Comprehension (ICPC'20)},
    doi          = {10.1145/3387904.3389258},
}

@inproceedings{Sulir2018,
	title        = {Integrating Runtime Values with Source Code to Facilitate Program Comprehension},
	author       = {Matúš Sulír},
	year         = 2018,
	booktitle      = {Proceedings of the 2018 IEEE International Conference on Software Maintenance and Evolution (ICSME'18)},
    doi          = {10.1109/ICSME.2018.00093},
}

@article{Suliretal2017,
	title        = {Source Code Documentation Generation Using Program Execution},
	author       = {Matúš Sulír and J. Porubän},
	year         = 2017,
	journal      = {Inf.},
    doi          = {10.3390/INFO8040148},

}

@inproceedings{Tanetal2023,
	title        = {Wait, wasn’t that code here before? Detecting Outdated Software Documentation},
	author       = {Wen Siang Tan and Markus Wagner and Christoph Treude},
	year         = 2023,
	booktitle      = {Proceedings of the 2023 IEEE International Conference on Software Maintenance and Evolution (ICSME'23)},
    doi          = {10.1109/ICSME58846.2023.00071},
}

@inproceedings{Wenetal2019,
	title        = {A Large-Scale Empirical Study on Code-Comment Inconsistencies},
	author       = {Fengcai Wen and Csaba Nagy and G. Bavota and Michele Lanza},
	year         = 2019,
	booktitle      = {Proceedings of the IEEE/ACM 27th International Conference on Program Comprehension (ICPC'19)},
    doi          = {10.1109/ICPC.2019.00019},
}

@inproceedings{Wesseletal2020,
	title        = {Effects of Adopting Code Review Bots on Pull Requests to OSS Projects},
	author       = {M. Wessel and Alexander Serebrenik and I. Wiese and Igor Steinmacher and M. Gerosa},
	year         = 2020,
	booktitle      = {Proceedings  of the 2020 IEEE International Conference on Software Maintenance and Evolution (ICSME'20)},
    doi          = {10.1109/ICSME46990.2020.00011},
}

@article{Zhouetal2020,
	title        = {Automatic Detection and Repair Recommendation of Directive Defects in Java API Documentation},
	author       = {Yu Zhou and Changzhi Wang and Xin Yan and Taolue Chen and Sebastiano Panichella and H. Gall},
	year         = 2020,
	journal      = {IEEE Transactions on Software Engineering},
    doi          = {10.1109/TSE.2018.2872971},
}

@inproceedings{Abukhalafetal2023,
	title        = {On Codex Prompt Engineering for OCL Generation: An Empirical Study},
	author       = {Seif Abukhalaf and Mohammad Hamdaqa and Foutse Khomh},
	year         = 2023,
	booktitle      = {Proceedings of the 20th {IEEE/ACM} International Conference on Mining Software Repositories (MSR'23)},
      doi          = {10.1109/MSR59073.2023.00033}
}

@inproceedings{Gaoetal2024,
	title        = {Search-Based LLMs for Code Optimization},
	author       = {Shuzheng Gao and Cuiyun Gao and Wenchao Gu and Michael R. Lyu},
	year         = 2024,
	booktitle      = {Proceedings of the 47th {IEEE/ACM} International Conference on Software Engineering (ICSE'24)},
  doi          = {10.1109/ICSE55347.2025.00021}
}

@inproceedings{Mondaletal2024,
	title        = {Enhancing User Interaction in ChatGPT: Characterizing and Consolidating Multiple Prompts for Issue Resolution},
	author       = {Saikat Mondal and Suborno Deb Bappon and C. Roy},
	year         = 2024,
	booktitle      = {Proceedings of the 21st International Conference on Mining Software Repositories (MSR'24)},
    doi       = {10.1145/3643991.3645085},
}

@inproceedings{Nashidetal2023,
	title        = {Retrieval-Based Prompt Selection for Code-Related Few-Shot Learning},
	author       = {Noor Nashid and Mifta Sintaha and A. Mesbah},
	year         = 2023,
	booktitle      = {Proceedings of the 2023 IEEE/ACM 45th International Conference on Software Engineering (ICSE'23)},
    doi = {10.1109/ICSE48619.2023.00205}
}

@inproceedings{Pengetal2023,
	title        = {Generative Type Inference for Python},
	author       = {Yun Peng and Chaozheng Wang and Wenxuan Wang and Cuiyun Gao and Michael R. Lyu},
	year         = 2023,
	booktitle      = {Proceedings of the 38th IEEE/ACM International Conference on Automated Software Engineering (ASE'23)},
    doi={10.1109/ASE56229.2023.00031}
}

@inproceedings{Renetal2023,
	title        = {From Misuse to Mastery: Enhancing Code Generation with Knowledge-Driven AI Chaining},
	author       = {Xiaoxue Ren and Xinyuan Ye and Dehai Zhao and Zhenchang Xing and Xiaohu Yang},
	year         = 2023,
	booktitle      = {Proceedings of the 38th IEEE/ACM International Conference on Automated Software Engineering (ASE'23)},
    doi={10.1109/ASE56229.2023.00143}
}

@inproceedings{Shinetal2023,
	title        = {Prompt Engineering or Fine-Tuning: An Empirical Assessment of LLMs for Code},
	author       = {Jiho Shin and Clark Tang and Tahmineh Mohati and Maleknaz Nayebi and Song Wang and Hadi Hemmati},
	year         = 2023,
	booktitle      = {Proceedings of the 2025 IEEE/ACM 22nd International Conference on Mining Software Repositories (MSR'25)},
    doi={10.1109/MSR66628.2025.00082}
}

@inproceedings{Shrivastavaetal2022,
	title        = {Repository-Level Prompt Generation for Large Language Models of Code},
	author       = {Disha Shrivastava and H. Larochelle and Daniel Tarlow},
	year         = 2022,
	booktitle      = {Proceedings of the International Conference on Machine Learning ({ICML}'23)},
    doi={10.5555/3618408.3619722}
}

@article{Wangetal2024,
	title        = {Agents in software engineering: survey, landscape, and vision},
	author       = {Yanlin Wang and Wanjun Zhong and Yanxian Huang and Ensheng Shi and Min Yang and Jiachi Chen and Hui Li and Yuchi Ma and Qianxiang Wang and Zibin Zheng},
	year         = 2024,
	journal      = {Automated Software Engineering},
    doi={10.1007/s10515-025-00544-2}
}

@inproceedings{Xiaetal2022,
	title        = {Automated Program Repair in the Era of Large Pre-trained Language Models},
	author       = {Chun Xia and Yuxiang Wei and Lingming Zhang},
	year         = 2023,
	booktitle      = {Proceedings of the 45th {IEEE/ACM} International Conference on Software Engineering (ICSE'23)},
    doi={10.1109/ICSE48619.2023.00129}
}

@inproceedings{Yeetal2023,
	title        = {ITER: Iterative Neural Repair for Multi-Location Patches},
	author       = {He Ye and Monperrus Martin},
	year         = 2023,
	booktitle      = {Proceedings of the 46th IEEE/ACM International Conference on Software Engineering (ICSE'23)},
    doi          = {10.1145/3597503.3623337},
}

@inproceedings{Zhangetal2024,
	title        = {Instruct or Interact? Exploring and Eliciting LLMs' Capability in Code Snippet Adaptation Through Prompt Engineering},
	author       = {Tanghaoran Zhang and Yue Yu and Xinjun Mao and Shangwen Wang and Kang Yang and Yao Lu and Zhang Zhang and Yuxin Zhao},
	year         = 2024,
	booktitle      = {Proceedings of the IEEE/ACM 47th International Conference on Software Engineering (ICSE'24)},
    doi={10.1109/ICSE55347.2025.00104}
}

@inproceedings{jimenez2024swebench,
title={{SWE}-bench: Can Language Models Resolve Real-world Github Issues?},
author={Carlos E Jimenez and John Yang and Alexander Wettig and Shunyu Yao and Kexin Pei and Ofir Press and Karthik R Narasimhan},
booktitle={Proceedings of the Twelfth International Conference on Learning Representations (ICLR'24)},
year={2024},
}

@inproceedings{zhang2024autocoderover,
author = {Zhang, Yuntong and Ruan, Haifeng and Fan, Zhiyu and Roychoudhury, Abhik},
title = {AutoCodeRover: Autonomous Program Improvement},
year = {2024},
doi = {10.1145/3650212.3680384},
booktitle = {Proceedings of the 33rd ACM SIGSOFT International Symposium on Software Testing and Analysis (ISSTA'24)},
pages = {1592--1604},
doi={10.1145/3650212.3680384}
}

@inproceedings{DBLP:conf/chi/Zamfirescu-Pereira23,
  author       = {J. D. Zamfirescu{-}Pereira and
                  Richmond Y. Wong and
                  Bjoern Hartmann and
                  Qian Yang},
  title        = {Why Johnny Can't Prompt: How Non-AI Experts Try (and Fail) to Design
                  {LLM} Prompts},
  booktitle    = {Proceedings of the 2023 {CHI} Conference on Human Factors in Computing
                  Systems ({CHI}'23)},
  pages        = {437:1--437:21},
  year         = {2023},
  doi          = {10.1145/3544548.3581388},
}

@inproceedings{DBLP:conf/icse/MastropaoloPGCSOB23,
  author       = {Antonio Mastropaolo and
                  Luca Pascarella and
                  Emanuela Guglielmi and
                  Matteo Ciniselli and
                  Simone Scalabrino and
                  Rocco Oliveto and
                  Gabriele Bavota},
  title        = {On the Robustness of Code Generation Techniques: An Empirical Study
                  on GitHub Copilot},
  booktitle    = {Proceedings of the 45th {IEEE/ACM} International Conference on Software Engineering({ICSE}'23)},
  pages        = {2149--2160},
  year         = {2023},
  doi          = {10.1109/ICSE48619.2023.00181}
}

@inproceedings{DBLP:conf/llm4code/RasnayakaWSI24,
  author       = {Sanka Rasnayaka and
                  Guanlin Wang and
                  Ridwan Shariffdeen and
                  Ganesh Neelakanta Iyer},
  title        = {An Empirical Study on Usage and Perceptions of LLMs in a Software
                  Engineering Project},
  booktitle    = {Proceedings of the 1st International Workshop on Large Language Models for Code (LLM4CODE'24)},
  pages        = {111--118},
  year         = {2024},
  doi          = {10.1145/3643795.3648379},
}

@article{lulla2026agentsmd,
      title={On the Impact of AGENTS.md Files on the Efficiency of AI Coding Agents}, 
      author={Jai Lal Lulla and Seyedmoein Mohsenimofidi and Matthias Galster and Jie M. Zhang and Sebastian Baltes and Christoph Treude},
      year={2026},
      eprint={2601.20404},
      archivePrefix={arXiv},
      primaryClass={cs.SE},
      url={https://arxiv.org/abs/2601.20404}, 
      doi= {10.48550/arXiv.2601.20404}
}

@article{gloaguen2026agentsmd,
      title={Evaluating AGENTS.md: Are Repository-Level Context Files Helpful for Coding Agents?}, 
      author={Thibaud Gloaguen and Niels Mündler and Mark Müller and Veselin Raychev and Martin Vechev},
      year={2026},
      eprint={2602.11988},
      archivePrefix={arXiv},
      url={https://arxiv.org/abs/2602.11988}, 
      doi={10.48550/arXiv.2602.11988
}
}

@article{li2025aidev,
      title={The Rise of AI Teammates in Software Engineering (SE) 3.0: How Autonomous Coding Agents Are Reshaping Software Engineering}, 
      author={Hao Li and Haoxiang Zhang and Ahmed E. Hassan},
      year={2025},
      eprint={2507.15003},
      archivePrefix={arXiv},
      doi={10.48550/arXiv.2507.15003}
}

@article{akhoroz2025conversationalaicodingassistant,
  author       = {Mehmet Akhoroz and
                  Caglar Yildirim},
  title        = {Conversational {AI} as a Coding Assistant: Understanding Programmers'
                  Interactions with and Expectations from Large Language Models for
                  Coding},
  year         = {2025},
  archivePrefix    = {arXiv},
  eprint       = {2503.16508},
  doi={10.48550/arXiv.2503.16508}
}

@article{Bubecketal2023,
author       = {S{\'{e}}bastien Bubeck and
                  Varun Chandrasekaran and
                  Ronen Eldan and
                  Johannes Gehrke and
                  Eric Horvitz and
                  Ece Kamar and
                  Peter Lee and
                  Yin Tat Lee and
                  Yuanzhi Li and
                  Scott M. Lundberg and
                  Harsha Nori and
                  Hamid Palangi and
                  Marco T{\'{u}}lio Ribeiro and
                  Yi Zhang},
  title        = {Sparks of Artificial General Intelligence: Early experiments with
                  {GPT-4}},
  doi          = {10.48550/ARXIV.2303.12712},
  archivePrefix    = {arXiv},
  eprint       = {2303.12712},
    year         = {2023}
}

@book{strauss1998basics,
  title={Basics of Qualitative Research: Techniques and Procedures for Developing Grounded Theory},
  author={Strauss, Anselm and Corbin, Juliet},
  year={1998},
  edition={2nd}
}

@book{auerbach2003qualitative,
  title={Qualitative Data: An Introduction to Coding and Analysis},
  author={Auerbach, Carl F. and Silverstein, Louise B.},
  year={2003},
  publisher={New York University Press},
  address={New York, NY},
  series={Qualitative Studies in Psychology},
  isbn={978-0814706954}
}

@book{saldana2021coding,
  title={The Coding Manual for Qualitative Researchers},
  author={Salda{\~n}a, Johnny},
  edition={4th},
  year={2021},
  publisher={SAGE Publications Ltd},
  address={London, UK},
  isbn={978-1529731743}
}

\appendix
\UseRawInputEncoding
\section{Appendix}\label{sec:appendix}

\subsection{Mapping of Manifest Headers to Taxonomy Labels}\label{app:label_mapping}

Table~\ref{tab:label_mapping} details the systematic mapping of original manual labels and representative headers found in the agent context files (\copilot, \claude, \codex) to the final categories used in our qualitative analysis.

\begin{table}[h!]
\centering
\caption{Taxonomy Mapping for Agent Context File Content}
\label{tab:label_mapping}
\scriptsize
\begin{tabular}{lll}
\hline
\textbf{Original Label} & \textbf{Example Header from Agent Context File} & \textbf{Final Merged Label} \\ \hline
Decision & Go/No-Go Decision Points & Project Management \\
Administration & Task Management & Project Management \\
Algorithms and Data Structures & Search Flow & Implementation Details \\
API \& Interface & Run API server & Build and Run \\
Architecture & Architecture & Architecture \\
Backend \& API & API Key (Development) & Architecture \\
Best Practices & Dashboard Implementation Best Practices & Implementation Details \\
Bibliography & References & Documentation and References \\
Build \& Test & Building and Running & Build and Run \\
Business Domain & The Business Logic Contract & System Overview \\
Business Logic & Business Needs & System Overview \\
CI/CD & CI/CD Pipeline & DevOps \\
AI Integration & Claude Code Configuration & AI Integration \\
Code Style & Code Style Guidelines & Implementation Details \\
Collaboration & Common Workflow Patterns & Development Process \\
Communication & Agent Communication Flow & Architecture \\
Component & Conversation Manager (\texttt{conversationManager.js}) & Architecture \\
Configuration & Environment Variables & Configuration and Environment \\
Context & Critical Context & System Overview \\
Data Management & Primary Data Sources & Architecture \\
Data Science and ML & Model breakdown display & Architecture \\
Data Structure & Data Structure Classes & Architecture \\
Database & Connect to database & Architecture \\
Dependencies and Libraries & Dependency Injection Setup & Architecture \\
Deployment and Operations & Production Deployment Details & DevOps \\
Development Commands & Build and Run Commands & Build and Run \\
Development Environment & Setup Docker Operations & Configuration and Environment \\
Development Timeline & Phase 2: Trust \& Transparency & Project Management \\
Development Workflow & Bug Fix Workflow & Development Process \\
Documentation & Key Documentation References & Documentation and References \\
Error Handling \& Debugging & Debugging \& Troubleshooting & Debugging \\
Example & Code Examples & Documentation and References \\
Features & Camera System & System Overview \\
Implementation Details & Important Technical Details & Implementation Details \\
Infrastructure Configuration & Ansible & Configuration and Environment \\
Integration & Event System Integration & Architecture \\
Limitations & Known Limitations & System Overview \\
Logging & Logging Infrastructure & Debugging \\
Memory Management & Memory Usage & Performance \\
Network and Communication & Routing and Request Handling & Architecture \\
Notes & Technical Notes & Documentation and References \\
Operating System & VeridianOS Overview & System Overview \\
Overview & Application Overview & System Overview \\
Performance \& QA & Performance and Monitoring & Performance \\
Project Management & Task Assessment Phase & Project Management \\
Project Operations & Sheet Operations & Implementation Details \\
Project Overview & Project Description & System Overview \\
Project Structure & Current Project Structure & Architecture \\
Project Workflow & Initiative Completion Flow & Development Process \\
Refactoring & Recent Improvements & Maintenance \\
Reminder & Auto-Push Reminder & Development Process \\
Security & Security Considerations & Security \\
Success Metrics & Success Criteria & Project Management \\
System Workflow & Execution Flow & Architecture \\
Task & Development TODOs & Project Management \\
Technology Stack & Framework Stack & Architecture \\
Tips & Development Tips & Development Process \\
Tools & Additional Tools & Development Process \\
UI/UX & User Experience & UI/UX \\
Version Control \& Git & Git and Development Workflow & Development Process \\
Work Progress & Current Progress & Project Management \\ \hline
\end{tabular}
\end{table}

\end{document}